\documentclass[a4paper,twocolumn,11pt,accepted=2026-04-18]{quantumarticle}
\pdfoutput=1
\usepackage{physics}
\usepackage{amsmath}
\usepackage{amssymb}
\usepackage{graphicx}
\usepackage{eczoo}
\usepackage{siunitx}
\usepackage[colorlinks=true, hyperindex, breaklinks, linkcolor=blue, urlcolor=blue, citecolor=blue]{hyperref}

\usepackage[capitalise]{cleveref}

\newcommand{\codepar}[1]{\ensuremath{[\![#1]\!]}}

\newcommand{\dqc}{Duke Quantum Center, Duke University, Durham, NC 27701, USA}
\newcommand{\physics}{Department of Physics, Duke University, Durham, NC 27708, USA}
\newcommand{\ece}{Department of Electrical and Computer Engineering, Duke University, Durham, NC 27708, USA}
\newcommand{\chem}{Department of Chemistry, Duke University, Durham, NC 27708, USA}

\begin{document}


\title{Hyper-optimized Quantum Lego Contraction Schedules}

\author{Balint Pato}
\email{balint.pato@duke.edu}
\affiliation{\dqc}
\affiliation{\ece}

\author{June Vanlerberghe}
\affiliation{\dqc}
\affiliation{\ece}

\author{Kenneth R. Brown}
\affiliation{\dqc}
\affiliation{\ece}
\affiliation{\physics}
\affiliation{\chem}


\date{2025-10-20}

\begin{abstract}
    Calculating the quantum weight enumerator polynomial (WEP) is a valuable tool for characterizing quantum error-correcting (QEC) codes, but it is computationally hard for large or complex codes. The Quantum LEGO (QL) framework provides a tensor network approach for WEP calculation, in some cases offering superpolynomial speedups over brute-force methods, provided the code exhibits area law entanglement, that a good QL layout is used, and an efficient tensor network contraction schedule is found. We analyze the performance of a hyper-optimized contraction schedule framework across QL layouts for diverse stabilizer code families. We find that the intermediate tensors in the QL networks for stabilizer WEPs are often highly sparse, invalidating the dense-tensor assumption of standard cost functions. To address this, we introduce an exact, polynomial-time Sparse Stabilizer Tensor (SST) cost function based on the rank of the parity check matrices for intermediate tensors. The SST cost function correlates perfectly with the true contraction cost, providing a significant advantage over the default cost function, which exhibits large uncertainty. Optimizing contraction schedules using the SST cost function yields substantial performance gains, achieving up to orders of magnitude improvement in actual contraction cost compared to using the dense tensor cost function. Furthermore, the precise cost estimation from the SST function offers an efficient metric to decide whether the QL-based WEP calculation is computationally superior to brute force for a given QL layout. These results, enabled by \mbox{PlanqTN}, a new open-source QL implementation, validate hyper-optimized contraction as a crucial technique for leveraging the QL framework to explore the QEC code design space.
\end{abstract}


\maketitle

\section{Introduction}


Quantum error correction (QEC) provides a path for creating reliable quantum computers from noisy components. Stabilizer codes \cite{gottesmanStabilizerCodesQuantum1997c} are the largest family of well-studied QEC codes and the basis of leading proposals for practical implementations.  Theoretically, these codes have only been exhaustively characterized up to distance 9 \cite{crossSmallBinaryStabilizer2025b} and explored broadly for topological codes \cite{bombinOptimalResourcesTopological2007, bombinTopologicalQuantumDistillation2006c, tillichQuantumLDPCCodes2009}, finite-rate QLDPC codes via algebraic constructions \cite{panteleevDegenerateQuantumLDPC2021a,wangDemonstrationLowoverheadQuantum2025,bravyiHighthresholdLowoverheadFaulttolerant2024}, concatenated codes \cite{yamasakiTimeEfficientConstantSpaceOverheadFaultTolerant2024a}, and holographic codes \cite{pastawskiHolographicQuantumErrorcorrecting2015a, caoApproximateBaconShorCode2021a,steinbergUniversalFaulttolerantLogic2025}. Yet the space of possible codes remains underexplored.

The quantum LEGO (QL) framework \cite{caoQuantumLegoBuilding2022} offers a universal way of constructing quantum error-correcting codes, which generalizes code concatenation, based on tensor networks. With the rising interest in machine learning, QL also might play a role as a language for artificial agents to explore the design space \cite{suDiscoveryOptimalQuantum2023}. The QL description of a QEC code can also speed up the calculation of the quantum weight enumerator polynomials (WEPs) \cite{shorQuantumAnalogMacWilliams1997c, caoQuantumWeightEnumerators2024} compared to exponential brute force methods. WEPs can help explore a rich set of characteristics of QEC codes, such as the code distance and behavior under coherent noise, and aid in decoding \cite{caoQuantumLegoExpansion2024a}. When the QL method is better than brute force is not entirely clear in practice, due to three contributing factors: first, a QEC code inherently has an entanglement structure, which, if too complex, can make brute force more advantageous than QL-based WEP calculation. Second, the QL layout for a QEC code is not unique, and there can be significant performance differences between layouts for the same code. Third, one must find an effective tensor network contraction schedule. Good contraction schedules are known for simpler cases, such as tree tensor networks corresponding to concatenated codes and loopy tree networks corresponding to holographic codes. However, finding an optimal contraction schedule in general is a \#P-hard problem \cite{dammComplexityTensorCalculus2002, valiantComplexityComputingPermanent1979}. Thus, a priori, for a general stabilizer code, it is unclear whether QL or brute-force is better for WEP calculation.

In this paper, we analyze the performance of a hyper-optimized contraction schedule framework \cite{grayHyperoptimizedTensorNetwork2021} for calculating WEPs of stabilizer code and QL layout families of varying complexity levels. This framework, implemented as the open source library Cotengra \cite{grayJcmgrayCotengra2025}, samples from the manifold of contraction schedules and provides the lowest-cost option. We find that the default cost function, which measures the number of operations during a full tensor network contraction, yields acceptable results, but it demonstrates large uncertainty in the resulting trees for the same code. It is designed for networks of dense tensors, and our sparse stabilizer code-specific cost function for weight enumerator tensor networks performs significantly better in a wide range of QL layouts, with up to an order of magnitude improvement. Beyond cost improvement, the precise knowledge of contraction cost reduces uncertainty and can also help decide whether the QL method is worthwhile or the brute force option is better. Our results utilize new quantum LEGO software tools, named PlanqTN \cite{planqtn_010}, which are available as an open-source Python library and a freely available web application.

The paper's structure is as follows. In \cref{sec:background} we introduce stabilizer codes, quantum weight enumerators, and the quantum LEGO formalism, including our extension to the language. The details of the QL layouts are described in \cref{sec:layouts}. We report our hyper-optimization cost function, the numerical findings for contraction costs, and some details of our software tools in \cref{sec:results}. Finally, we conclude with a summary of our results and our outlook in \cref{sec:conclusions}.

\section{Background}\label{sec:background}

In this section, we introduce the stabilizer formalism, the notion of scalar and tensor weight enumerator polynomials for stabilizer codes, and the quantum LEGO formalism, along with its application in calculating weight enumerators. While the quantum LEGO framework applies in full generality to qudits and complex noise models, for simplicity and our purposes, we restrict the discussion to qubits, stabilizer codes, and the reduced weight enumerators \cite{caoQuantumLegoExpansion2024a}. The $n$-qubit Hilbert space is denoted $\mathcal{H}_n=\otimes_n\mathbb{C}^2$.

\subsection{Stabilizer codes}

Most, if not all, practical quantum error correcting codes belong to the family of \textit{stabilizer codes} \cite{gottesmanStabilizerCodesQuantum1997c}. We denote the $n$-qubit Pauli group with $P_n$. An \codepar{n,k,d} qubit stabilizer code is a $k$-qubit subspace of $n$ physical qubits, with minimum distance $d$, defined by the shared +1 eigenspace of $n-k$ operators in $\mathcal{P}_n$ called the \textit{stabilizer generators}. The stabilizer generators span the stabilizer group $S$, which is an Abelian group of order $2^{n-k}$.

A set of stabilizer generators can be represented as the \textit{parity check matrix} (PCM) of a stabilizer code, where each row is a stabilizer generator in symplectic form:

\begin{align}
    H=\left(
    \begin{array}{c|c}
        H_X & H_Z
    \end{array}
    \right),
\end{align}

\noindent where $H_X$ ($H_Z$) is an $n-k \times n$ binary matrix that corresponds to the $X$ ($Z$)-part of each stabilizer generator qubit-by-qubit, meaning that $0|0$ is $I$, $1|0$ is $X$, $0|1$ is $Z$ and $1|1$ is the $Y$ operator. For example, the \codepar{4,1,2} code ($d=2$ rotated surface code) has stabilizer generators $\langle XXXX,IIZZ, ZZII\rangle$ and will have symplectic parity check matrix:
$$
    \left(
    \begin{array}{cccc|cccc}
            1 & 1 & 1 & 1 & 0 & 0 & 0 & 0 \\
            0 & 0 & 0 & 0 & 1 & 1 & 0 & 0 \\
            0 & 0 & 0 & 0 & 0 & 0 & 1 & 1 \\
        \end{array}
    \right)
$$

A PCM $H$ can contain an overcomplete set of stabilizers, in which case the row rank, denoted $rank(H)$, is smaller than the number of rows. The rowspace of the PCM $H$ of a stabilizer code, denoted $row(H)$ equals to $S$, the stabilizer group. Given a set of qubits indexed by $J \subset \{1, \ldots,n\}$, we can talk about a submatrix of the PCM $H$ (a sub-PCM), denoted $H|_J$, which due to the symplectic form will contain $2|J|$ columns to represent the $X$ and $Z$ parts.

We will denote with $\Pi_C = \frac{1}{2^{n-k}}\sum_{M \in S} M $ the projector to the code space of a stabilizer code $C$ with stabilizer group $S$.

Logical Pauli operators are in $\mathcal{N}_{P_n}(S)$, the normalizer of the stabilizer group within $P_n$, with elements of $S$ corresponding to the trivial logical operator, the identity. The \textit{encoding map} of a code $C$ is the isometry $V_C: \mathcal{H}_k \rightarrow \mathcal{H}_n$ mapping the $k$-qubit logical states to the $n$-qubit physical states:

\begin{align}
    \sum_{l_j \in \{0,1\}} V_{l_1 \ldots l_{n+k}} \ketbra{l_1 \ldots l_n}{l_{n+1}\ldots l_{n+k}},
\end{align}

\noindent where we denoted the individual qubit degrees of freedom with $l_j$, the first $n$ being the physical qubits, and the last $k$ being the logical qubits. With this description, it is apparent that the coefficients $V_{l_1, \ldots, l_{n+k}}$ can be represented by a tensor with $n+k$ \textit{legs} (hence the $l$ notation for the indices), meaning a rank-$n+k$ tensor. The \textit{encoding tensor} of a stabilizer code will be the unnormalized Choi state:

\begin{align}
    \ket{V}\! =\!\sum_{l_j \in \{0,1\}}\!V_{l_1 \ldots l_{n+k}} \ket{l_1 \ldots l_{n+k}} \label{eq: encoding tensor}.
\end{align}

For an operator $E \in P_n$, the \textit{weight of $E$} counts the number of qubits where $E$ acts non-trivially, and will be denoted $0 \leq |E| \leq n$. The stabilizer code's min distance $d$ is the minimum weight over the logical Pauli operators $d=\min\{|E|: E \in C_{P_n}(S)\}$. Calculating the distance in general is an NP-hard problem, and while faster, heuristic methods are available \cite{pryadkoQDistRndGAPPackage2022a}. One way to determine the distance exactly is via the calculation of quantum weight enumerator polynomials.

\subsection{Weight enumerator polynomials}

Shor and Laflamme generalized the notion of weight enumerator polynomials (WEP) to quantum observables \cite{shorQuantumAnalogMacWilliams1997c} by defining two types of WEPs for two observables $O_1$ and $O_2$ over an arbitrary trace-orthonormal \textit{error basis}, $\mathcal{E}$:

\begin{align}
    A(z;O_1; O_2) & = \sum_{E \in \mathcal{E}} \frac{\Tr [E^\dagger O_1 ] \Tr[ E O_2 ] z^{|E|}}{\Tr O_1\Tr O_2} \label{eq: shor laflamme A} \\
    B(z;O_1; O_2) & = \sum_{E \in \mathcal{E}} \frac{ \Tr [E^\dagger O_1  E O_2 ] z^{|E|}}{\Tr O_1 O_2} \label{eq: shor laflamme B}
\end{align}

For a qubit stabilizer code $C$, with code space projector $\Pi_C$, the observables in \cref{eq: shor laflamme A,eq: shor laflamme B} become $O_1=O_2=\Pi_C$ in \cref{eq: code enum A,eq: code enum B} and the error basis $\mathcal{E}$ becomes $\mathcal{P}_n$. As $P^\dagger=P$ for all Paulis, the formulas become simpler, the \textit{stabilizer weight enumerator polynomial} ($A$-type) and the \textit{normalizer weight enumerator polynomial} ($B$-type) for qubit stabilizer code $C$ are:

\begin{align}
    A_C(z;\Pi_C) & =  \frac{1}{4^k} \sum_{P \in \mathcal{P}_n} (\Tr[P \Pi_C ])^2 z^{|P|} \label{eq: code enum A}     \\
    B_C(z;\Pi_C) & = \frac{1}{2^k} \sum_{P \in \mathcal{P}_n} \Tr[P \Pi_C  P \Pi_C ] z^{|P|} \label{eq: code enum B}
\end{align}

From this form, it is clear why these polynomials enumerate operator weights - since

$$\Tr(P\Pi_C) = \begin{cases}\frac{2^n}{2^{n-k}}=2^k, P \in S\\0, P \notin S\end{cases},$$

the coefficient of $z^w$ will be the number of weight-$w$ stabilizers. Similarly, if $P$ is in the normalizer of $S$, then $P^\dagger \Pi_C P=\Pi_C$, resulting in:

$$
    \Tr[P \Pi_C  P \Pi_C ]=\begin{cases}
        2^k, P \in \mathcal{N}_{\mathcal{P}_n}(S) \\
        0, P \notin \mathcal{N}_{\mathcal{P}_n}(S)
    \end{cases}.
$$

Notably, the minimum distance of an \codepar{n,k,d} code is the minimum power $d$ of $B_C(z)-A_C(z)$. Furthermore, it is sufficient to calculate $A_C(z)$, as $B_C(z)$ can be expressed using the quantum MacWilliams identities \cite{shorQuantumAnalogMacWilliams1997c}:

\begin{align}
    B_C(z;\Pi_C) & = \frac{1}{2^{n-k}}(1+3z)^n A(\frac{1-z}{1+3z};\Pi_C)
\end{align}

In the completely general treatment by Cao and Lackey \cite{caoQuantumWeightEnumerators2024}, the above weight enumerator polynomials are called the \textit{scalar} weight enumerator polynomials, as they are real-valued polynomials, elements of $\mathbb{R}[z]$, as opposed to the more general tensor WEPs, which we will introduce now.

In the next section, we will explore the quantum LEGO framework and how it enables the creation of new codes by building a tensor network from encoded tensors of smaller codes. These networks will be helpful to calculate other tensor types, namely \textit{tensor weight enumerator polynomials}, a more general form of WEPs that allow for calculating larger WEPs through contraction. While there are more general forms of complete and multivariate polynomials capturing more complex noise models and calculations, we will only focus on \textit{reduced} tensor weight enumerator polynomials \cite{caoQuantumWeightEnumerators2024}. This is warranted, as our discussion is focused on stabilizer codes, which always have diagonal tensor weight enumerators \cite{caoQuantumWeightEnumerators2024}.

To define a tensor weight enumerator polynomial, we leave a set of qubits $J \subset \{1, \ldots,n\}$ ``open'', which will correspond to grouping the stabilizers by operators on these qubits and aggregating the weight distribution within these groups. More formally, denoting $m = |J|$, for each Pauli operator $P \in \mathcal{P}_m$ we calculate a scalar WEP on the remaining indices, $J^C=\{1,\ldots,n\}\setminus J$ across all stabilizers that have $P$ on indices $J$. We store the results in a $4^m$ dimensional object, which can be represented as a rank-$m$ tensor with 4-dimensional legs, that belongs to a $\mathbb{R} \otimes V$, with $V=span_\mathbb{R}\{e_E: E \in \mathcal{P}_m\}$, where $\{e_E\}$ are orthonormal basis elements of a $4^m$-dimensional vector space. Thus, the formal definition for a tensor WEP for a code $C$ with codespace projector $\Pi_C$ with open legs $J \in \{1,\ldots,n\}$ is:

\begin{align}
    4^k A^{(J)}(z;\Pi_C) & =\!\!\!\!\!\!\sum_{\substack{E \in \mathcal{P}_{m}                             \\ F \in \mathcal{P}_{n-m}}}\!\!\!\!\!(\Tr[(E \otimes_J F )\Pi_C])^2 z^{|F|} e_E  \nonumber \\
                         & =  \sum_{E \in \mathcal{P}_{m}} A^{(J)}_E(z;\Pi_C)  e_E \label{eq: tensor wep}
\end{align}

\noindent where the $(E \otimes_J F)$ operator is a tensor product of single-qubit Pauli operators that agree with $E$ on the qubits in the set $J$, and agree with $F$ on $J^C$ and we denoted the scalar WEP corresponding to the element $E$ with $A^{(J)}_E(z;\Pi_C)$. We can now see that for $J=\emptyset$, the simplest special case of tensor weight enumerator polynomials is the scalar WEP, which considers the weight of all stabilizers in the stabilizer group. When $|J|=1$, \cref{eq: tensor wep} corresponds to a four-dimensional vector enumerator polynomial. A less abstract representation of a tensor WEP would be a sparse dictionary, with keys being the $E$ Paulis, and values being the corresponding scalar weight enumerators.

\subsection{Quantum LEGO formalism}

The Quantum LEGO \cite{caoQuantumLegoBuilding2022} framework is a tensor network description of stabilizer codes and has three major features: constructing a new stabilizer code from smaller codes alongside calculating its PCM, tracking logical operators through operator pushing, and finally calculating weight enumerator polynomials. In this paper, we will not focus on tracking logical operators; instead, we will focus on the code construction and WEP calculation aspects of the framework.

\subsubsection{PlanqTN implementation of quantum LEGO}

\begin{figure}[htbp!]
    \centering
    \includegraphics[width=0.9\linewidth]{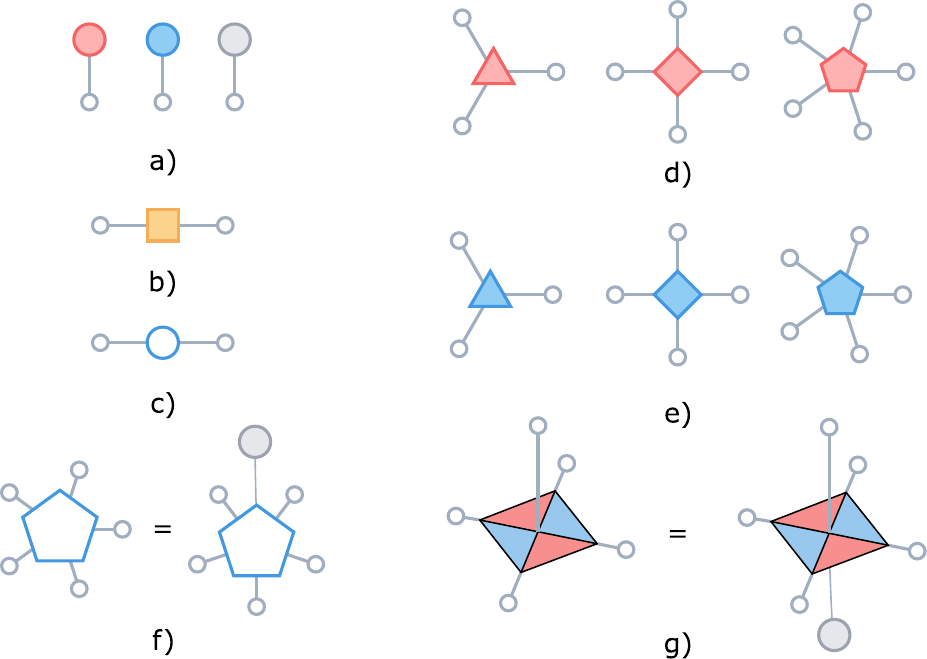}
    \caption{Quantum LEGO building blocks used in the paper. Stoppers (a) are single-qubit states, stabilized by Pauli-$X$ (red) or Pauli-$Z$ (blue) operators. The identity stopper (gray) is the ``free qubit'', which is only stabilized by the $I$ operator, and thus is a subspace LEGO. The Hadamard LEGO (b) is stabilized by $\langle XZ, ZX\rangle$. The identity LEGO (c) or Bell-state is stabilized by $\langle XX, ZZ\rangle$. The 3, 4, and 5-legged examples of bit-flip-code d) (phase-flip-code e)) encoding tensors stabilized by all weight-two $ZZ$ ($XX$) operators and the $X^{\otimes n}$ ($Z^{\otimes n}$) operator. The subspace LEGO of the \codepar{5,1,3} code stabilized by $\langle XZZXI,IXZZX,XIXZZ,ZXIXZ\rangle$ (f) equals the \codepar{6,0,4} perfect encoding tensor traced on the logical leg with an identity stopper. Similarly, the \codepar{5,1,2} code stabilized by $\langle XXXXI,ZZZZI,XXIIX,ZIZIZ\rangle$ (g) equals the \codepar{6,0,3} encoding tensor for the \codepar{4,2,2} code.}
    \label{fig: building blocks}
\end{figure}

All the quantum LEGO figures in this paper are prepared using the PlanqTN Studio software \cite{planqtn_010}, and the algorithms are implemented as part of version 0.2.0. We collected the LEGOs used in this paper in \cref{fig: building blocks} and provided their stabilizers in the description. The PlanqTN visual language closely follows the original QL definitions, with slight changes in colors and an extension called the ``identity stopper''. The identity stopper's PCM is $I$, the single-qubit identity operator, representing the free qubit. When traced on a leg with a tensor, the only surviving stabilizers are those that have $I$ on that leg. Thus, the identity stopper naturally serves to hide logical legs, creating a diagram that represents a non-trivial subspace, rather than a state (i.e., a zero-dimensional subspace).

\subsubsection{Quantum LEGO for code construction} \label{sssec: QL PCM}

For stabilizer codes, the encoding tensor \cref{eq: encoding tensor} is a stabilizer state. This state contains the logical legs as well as the physical legs. To continue the example of the \codepar{4,1,2}, code, choosing the logical representatives $\bar{X}=XXXX,\bar{Z}=ZZZZ$, we can define the PCM of the code's \codepar{5,0,2} encoding tensor as:

$$
    \begin{array}{c}
        \texttt{XXXXI} \\
        \texttt{ZZIII} \\
        \texttt{IIZZI} \\
        \texttt{XXXXX} \\
        \texttt{ZZZZZ}.
    \end{array}
$$

To construct new stabilizer codes in quantum LEGO, pairs of qubits from smaller codes' encoding tensors (or subspace legos) are traced together - we take the product of the Bell state and the two tensors on the given legs. For two encoding tensors $\ket{V_A}, \ket{V_B}$, with qubit indices to trace $a$ and $b$, the trace operation is formally as below:

\begin{align}
    \ket{V_A} \land_{a,b} \ket{V_B} & = \bra{\phi^+}_{ab} \ket{V_A} \otimes \ket{V_B}                                     \\
    \bra{\phi^+}_{ab}               & =\frac{\bra{0}_a\otimes \bra{0}_b + \bra{1}_a\otimes \bra{1}_b}{\sqrt{2}} \nonumber
\end{align}

Here, we describe an efficient algorithm for calculating the final encoding tensor of stabilizer codes, utilizing symplectic parity check matrices.

We follow the method described by Cao and Lackey \cite{caoApproximateBaconShorCode2021a,caoQuantumLegoBuilding2022}. Calculating the stabilizers for the full network (or any subgraphs) is polynomial, as tracing only requires Gauss elimination and elementary row operations, which we will describe now.

All stabilizer code tracing can be simplified to two operations: tensoring PCMs and self-tracing two legs together. When tracing two legs of two separate codes, first we tensor them together into a single code and then apply self-trace. Tracing multiple pairs of legs is equivalent to doing the traces separately. Elements of two independent subspaces can be described as the elements in the tensor product of the two subspaces. Thus, the tensor product is used to describe stabilizers of disjoint connected components of tensor networks. We can calculate the PCM of the joint $A \otimes B$ subspace of the subspaces $A$ and $B$, with PCMs $H_A$ and $H_B$, respectively, by taking the block sum of the two PCMs:

\begin{align}
    H_{A \otimes B} & = H_A \oplus H_B \\
                    & = \left(
    \begin{array}{cc|cc}
            (H_A)_X & 0       & (H_A)_Z & 0       \\
            0       & (H_B)_X & 0       & (H_B)_Z
        \end{array}
    \right) \nonumber
\end{align}

Self-tracing physically corresponds to measuring the two qubits in the Bell-basis (an $XX$ and a $ZZ$ measurement) and projecting to the $(+1, +1)$ measurement outcome. In other words, the stabilizers that will survive are those that commute with both $XX$ and $ZZ$ on the traced qubits, and these are the ones that have the exact same operators there. To find the generators for these stabilizers, we first execute a Gauss elimination limited to these four columns (the $X$ and $Z$ columns for each leg). Then, we have either one or zero elements in each column equal to 1. Denote the ``pivot row index'' - the first non-zero row for these columns by $x_1, x_2, z_1, z_2$. If a column is all zero, the pivot row index is set to -1. When $x_1=x_2$ ($z_1=z_2$), all generators will commute with the $ZZ$ ($XX$) measurement, and no changes are needed. Otherwise, if $x_1=-1$, then remove the row $x_2$ (similarly for the opposite case, and similarly to $z_1$ and $z_2$), as this corresponds to a case where no stabilizer containing that generator will commute with $ZZ$ (or $XX$ in the $z_i$'s case). The final case is when neither $x_1$ nor $x_2$ is equal to -1, but they are not equal to each other. Then, add $x_1$ to the $x_0$ row and remove $x_1$ (similarly, do the same for $ z_1$ and $ z_2$). At this point, all the stabilizers generated by the surviving generators commute with $XX$ and $ZZ$ on the two traced qubits. We can then remove the four columns corresponding to these two qubits. To ensure that the matrix is of full rank, we execute a final Gauss elimination on all columns and remove all potential all-zero rows. This algorithm has polynomial complexity and will serve as the basis for our cost function, which will later determine the cost of a contraction schedule.

\subsubsection{Quantum LEGO for weight enumerator polynomials} \label{sssec: QL WEP}

Surprisingly, a tensor network of stabilizer codes can be used to calculate the weight enumerator polynomial of the traced code if we know the WEP for the individual LEGOs. To put the correspondence more formally, tracing two open legs in a tensor weight enumerator is the same as calculating the WEP of the corresponding LEGO network after tracing the corresponding two legs, as per Theorem 2 in    \cite{caoQuantumLegoExpansion2024a}. The brute force method for calculating a WEP requires $O(2^{n-k})$ operations, as the naive implementation simply enumerates the elements of the stabilizer group. For certain QEC codes with a good QL layout and efficient contraction schedule, this method can outperform the brute force method. To execute this calculation, we create a new tensor network, now with tensor weight enumerator polynomials for each of the LEGOs, with open legs for all non-dangling legs. The first set of tensor WEPs is calculated via brute force. Then, for a given contraction schedule, we trace each connection in the network \cite{caoQuantumLegoExpansion2024a}. We will now describe how to contract a network of tensor weight enumerators.

Similar to the parity check matrix method, the tensor product describes the WEP of disconnected components. We can also describe the trace as a tensor product and a self-trace. Calculating the tensor product of two tensor enumerators $A^{(J_1)}(z;\Pi_{C_1}) \otimes A^{(J_2)}(z;\Pi_{C_2})$ with $J_1=\{1,\ldots,n\}, J_2=\{1,\ldots,m\}$ is straightforward. The basis elements will be the tensor product of the basis elements. For each tensored basis element, the corresponding WEP will be the product of the two scalar weight enumerator polynomials:
\begin{align}
    \sum_{\substack{E_1 \in \mathcal{P}_n \\ E_2 \in \mathcal{P}_m}} A_{E_1}^{(J_1)}(z;\Pi_{C_1}) A_{E_2}^{(J_2)}(z;\Pi_{C_2}) e_{E_1} \otimes e_{E_2}
\end{align}

For a self-trace, we sum over elements of the tensor where the indices are equal, and remove the traced indices.

\begin{align}
    \sum_{a_1,a_2,\ldots a_k \in \mathcal{P}_k} & A^{a_1,a_1, a_2,a_2\ldots,a_k, a_k,b_1,b_2,\ldots,b_n}(z) \nonumber \\
    =                                           & A^{b_1,b_2,\ldots,b_n}(z) \label{eq: wep-self-trace}
\end{align}

\subsection{Hyper-optimized tensor network contraction schedules}

Tensor networks are a widely used tool in quantum computing as well as in other scientific areas. The problems that can benefit are those with a computational task involving a large tensor that can be decomposed into a sparse network of small-rank tensors, thus making the computation more tractable. A common way of executing calculations on the full network is via \textit{tensor network contraction}, which can be thought of as matrix multiplications on the matching indices. In quantum computing and many-body physics, for a few network topologies (e.g., trees, grid layouts in 2D, 3D, loopy trees on a hyperbolic tiling) \cite{shiClassicalSimulationQuantum2006,valiantHolographicAlgorithms2008,mahdaviAdvancesQuantumComputation2009a,dennyAlgebraicallyContractibleTopological2011,caoQuantumLegoExpansion2024a}, there are known strategies to contract tensor networks to calculate expectation values. However, for an arbitrary graph, it is in general a \#P-problem to find an optimal sequence. It can result in orders of magnitude differences in performance, surprisingly even for grid layouts. Gray and Kourtis introduced a family of algorithms to construct hyper-optimized contraction schedules for arbitrary networks \cite{grayHyperoptimizedTensorNetwork2021}, which they also implemented as an open source software, under the name Cotengra \cite{grayJcmgrayCotengra2025}. On a wide variety of tensor networks, this method outperformed traditional methods by orders of magnitude in terms of cost of contraction. We will now briefly introduce the ideas behind Cotengra.

A tensor network is represented as an undirected graph $G=(V,E)$ with the set of vertices $V$ corresponding to tensors, and edges $E$, corresponding to traces between legs of tensors. While Gray and Kourtis's framework allows for hyperedges, we don't consider them here, and we will assume that all edges are graph-like edges. We note that the phase-flip and bit-flip code LEGOs, which also correspond to the Z and $X$-spiders in ZX-calculus, can be considered hyperedges. Still, for this paper, we will consider them as tensors and will not decompose them into pairwise edges. We will also assume that the network is fully connected and does not have self-loops initially.

\begin{figure}[htbp!]
    \centering
    \includegraphics[width=.7\linewidth]{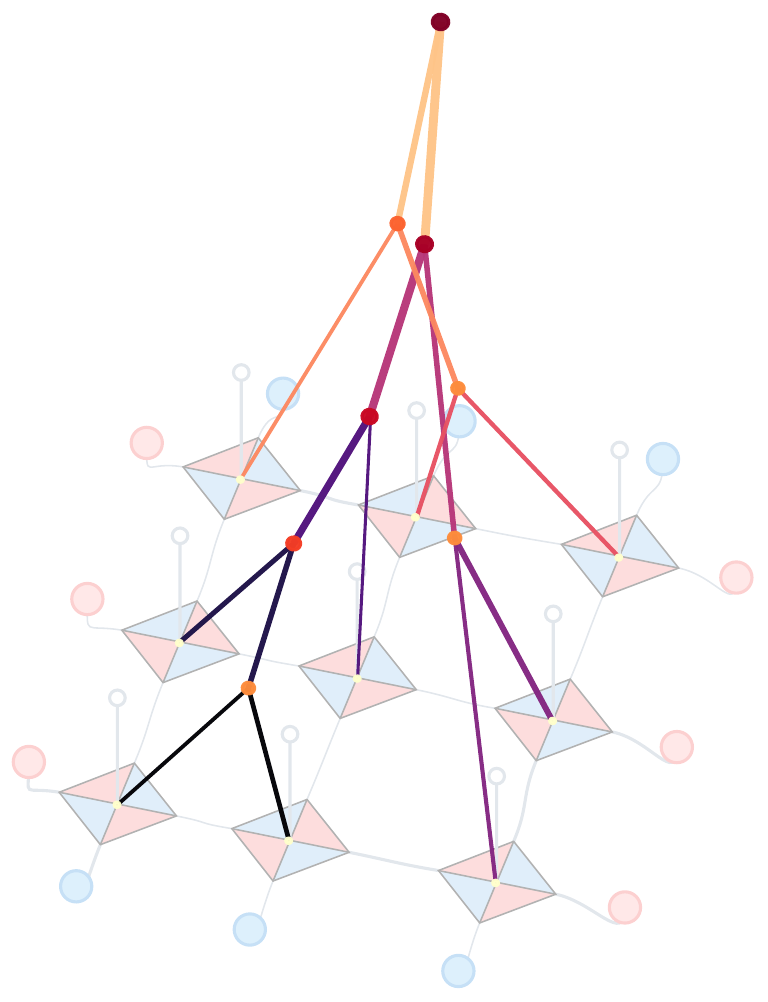}
    \caption{Cotengra contraction tree overlayed over a 3x3 rotated surface code quantum LEGO network. The $X$ and $Z$-stoppers are conjoined first with the \codepar{5,1,2} subspace LEGOs, and thus the Cotengra contraction only considers the resulting 9 qubit-wise tensors.}
    \label{fig:contraction-tree}
\end{figure}

\textit{Tensor network contraction} is represented as a sequence of vertex contractions in $G$. This sequence can be represented as a binary tree $B=(V_B,E_B)$, with vertices $V_B$ and edges $E_B$. The root represents the fully contracted network, leaves are the original vertices $V$ from $G$, and vertices $V_B$ of the tree represent vertex contractions in $G$. Each vertex contraction assumes contraction on all edges between the two components. For example, see \cref{fig:contraction-tree} for the 3x3 rotated surface code LEGO network overlaid with the contraction tree, as plotted by Cotengra. The four bottom left LEGOs are contracted through three contractions, the first two vertex contractions containing one edge each, while the third contraction contains two edges from the original graph $G$.

A contraction tree $B$ can then be evaluated for the cost of contraction in space or time. In this paper, we focus on the contraction cost in time.

To evaluate the time requirement, the \textit{contraction cost} is calculated, which is based on the total \textit{vertex congestion} \cite{ogormanParameterizationTensorNetwork2019}, that for each vertex contraction takes the product of the total number of elements in the left $l(v)$ and right $r(v)$ intermediate child tensors of $v$:
\begin{align}
    C(B) = \sum_{v \in V_B \setminus V} 4^{|u(l(v))|+|u(r(v))|}\ ,
\end{align}
where $u(v)$ is the set of uncontracted legs in the intermediate tensor $v$. Again, a core implicit assumption in these prior works is that the tensors are stored and operated on using a dense representation.

Now that we have a cost function, we can optimize trees using it; the time-optimal tree minimizes $C(B)$. Gray and Kourtis evaluated multiple strategies for optimizing the contraction tree cost. We will use three of them, namely, \texttt{Optimal}, \texttt{Hyper-Greedy}, and \texttt{Hyper-Par}.

The \texttt{Optimal} strategy evaluates all possible trees and picks the optimal one. Due to the exhaustive search, this strategy is only available for small networks.

The \texttt{Hyper-Greedy} strategy evaluates the tree bottom up, and greedily decides which two tensors to contract next. For two tensors $T_i, T_j$, that would result in a contracted tensor $T_k$, the probability of selecting them is:

\begin{align}
    p(T_i,T_j)  \propto e^{\left( \frac{\alpha(size(T_i) + size(T_j))-\frac{size(T_k)}{\alpha}}{\tau}\right)},
    \label{eq: greedy equation}
\end{align}

\noindent where hyperparameters $\alpha$ and $\tau$ are then used by a Bayesian optimizer to drive the sampling of trees towards the minimal cost function. The probability is thus weighted according to a Boltzmann distribution with temperature parameter $\tau$, which controls how broad the exploration should be. The exponentiated cost function is the change in tensor sizes at $\alpha=1$ and the new tensor's size at $\alpha=0$.

The \texttt{Hyper-Par} contraction strategy instead views the creation of a contraction tree as a hypergraph partitioning problem. It starts from the root of the tree corresponding to the whole graph, then begins to split it into $k$ partitions recursively using KaHyPar \cite{KaHyPar1,KaHyPar2} to generate partitions accordingly to an imbalance parameter $\epsilon$, so that $\forall i \in[1,k]:|V_i| \leq (1+\epsilon) |V|/k$. Then, amongst the $k$ partitions, if a sub-tree is small ($\leq 10$ nodes) it is contracted with the fast \texttt{Hyper-Greedy} strategy, otherwise if it's below the optimal contraction cost it will be contracted with the \texttt{Optimal} strategy, or, recursively, the \texttt{Hyper-Par} will be used to calculate the contraction tree. The Bayesian optimizer is then used to execute a hyper-optimization by tuning the hyperparameters $k$ and $\epsilon$.

Both the \texttt{Hyper-Greedy} and \texttt{Hyper-Par} strategies are inherently random, and the number of trials can be considered another hyperparameter. Cotengra also offers a time-based cutoff for trials for practical use cases.

Finally, we emphasize that the cost function is central to the Cotengra strategies, and therefore, the implicit dense tensor assumption plays a crucial role. As we will see in the following sections, the quantum LEGO layouts we examined for stabilizer codes yield highly sparse tensors. Thus, a cost function representing a more accurate picture of contraction costs can provide a significant improvement over the default cost functions.

\section{Quantum LEGO layouts}\label{sec:layouts}

In this section, we describe the details of the tensor network layouts for the quantum codes studied in this paper. We picked an example for each of the classes as defined by Cao and Lackey \cite{caoQuantumLegoExpansion2024a}, namely, concatenated tree layouts, holographic codes, surface codes, and two universal layouts that can be calculated for any stabilizer code: the measurement state preparation and Tanner networks.

\subsection{Concatenated code, tree network}

Given two codes $C_1$, $C_2$ (an \textit{inner} and an \textit{outer} code, respectively) with parameters $\codepar{n_1,1,d_1}$ and $\codepar{n_2,1,d_2}$ we can encode the physical qubits of $C_2$ into the code of $C_1$, resulting in a \codepar{n_1 n_2, 1, d_1 d_2} code \cite{gottesmanStabilizerCodesQuantum1997c}. This construction is called \textit{code concatenation}, and it is naturally represented in the QL framework by connecting the physical legs of $C_2$ with the logical legs of $C_1$, resulting in a tree structure, in which each level of the tree corresponds to a level of code concatenation. We note that codes with more than one logical qubit can also be concatenated; however, the result may not be a tree, but rather a more densely connected graph.

\begin{figure}
    \centering
    \includegraphics[width=0.7\linewidth]{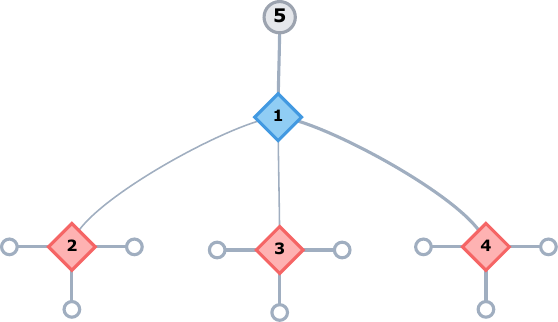}
    \caption{Concatenated layout for Shor's code. The blue LEGO is the encoding tensor of the $d=3$ phase-flip code with a single logical leg, while the red LEGOs are the $d=3$ bitflip codes with a single logical leg.}
    \label{fig:conc-shor}
\end{figure}

In our case, we concatenate the bitflip and phase-flip repetition codes in alternating layers as shown in \cref{fig:conc-shor}. The first two layers for $d=3$ repetition codes result in Shor's code \cite{shorFaulttolerantQuantumComputation1996}. We construct two different generalized Shor's code families with increasing qubits and distance. The first one is the typical one, resulting in the same tree but with higher distance repetition codes. The second one is using the same $d=3$ repetition codes but has more concatenation layers.

Tree tensor networks (TTN) can be contracted in $O(n)$ time in the general case, or, if all nodes are identical, then in $O(\log n)$ time, by an explicit, simple layer-by-layer algorithm \cite{caoQuantumLegoExpansion2024a}. This leads to super-polynomial speedups compared to brute force using the QL contraction.

\subsection{Holographic code}

\begin{figure}[!htbp]
    \centering
    \includegraphics[width=0.9\linewidth]{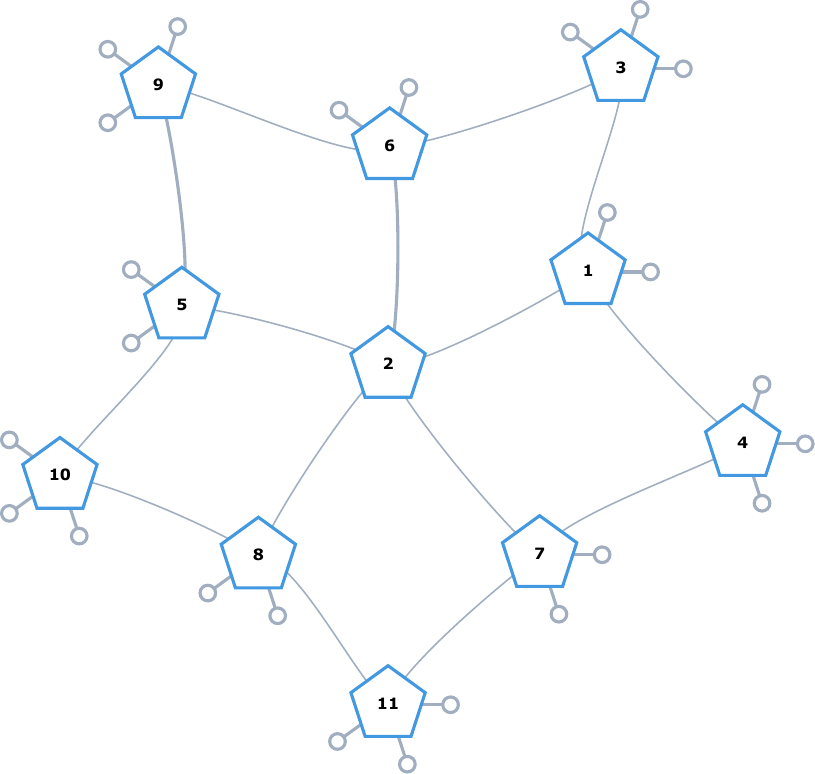}
    \caption{The HaPPY \cite{pastawskiHolographicQuantumErrorcorrecting2015a} code with perfect encoding tensors of the \codepar{5,1,3} code. The logical legs are omitted, making these LEGOs subspace LEGOs.}
    \label{fig:HaPPy}
\end{figure}

A slightly more complex class of layouts is \textit{trees with loops}, which occur in certain holographic code families. The HaPPY code \cite{pastawskiHolographicQuantumErrorcorrecting2015a} is a canonical example, where physical legs of encoding tensors of the \codepar{5,1,3} code are glued together in a hyperbolic tiling space. This translates to one logical leg per tensor across the network in both the bulk and the boundary tensors, and leaves physical legs only on the boundary. In \cref{fig:HaPPy} we display a member of this family, using subspace LEGOs (with logical legs omitted).

Cao, Gullans, Lackey and Wang estimate the upper bound on contraction cost of these codes to be $O(n^{1+\alpha}), \alpha>0$, where $\alpha$ is dependent on the exact hyperbolic tesselation \cite{caoQuantumLegoExpansion2024a}, resulting in super-polynomial speedup compared to brute force using the QL contraction.

\subsection{Two-dimensional grid network}

The rotated surface code family can be implemented as a rectangular two-dimensional grid of \codepar{6,0,3} encoding tensors \cite{caoQuantumLegoBuilding2022} and $X$ and $Z$-stoppers on the boundaries corresponding to $X$ and $Z$ boundaries. In \cref{fig:contraction-tree}, we show an example 3x3 rotated surface code layout, where the bottom logical leg of each \codepar{6,0,3} tensor is hidden, making each LEGO into a \codepar{5,1,2} subspace LEGO.

For a $D$-dimensional Euclidean lattice, the upper bound for the contraction cost is $O(n \exp(n^{1-1/D}))$ as per \cite{caoQuantumLegoExpansion2024a}.

\subsection{Measurement state preparation and Tanner networks}

As described by Cao and Lackey \cite{caoQuantumLegoBuilding2022}, there are universally available representations for stabilizer codes; our examples will be based on their measurement state preparation (MSP) circuit and Tanner graph constructions.

\begin{figure}[!htbp]
    \centering
    \includegraphics[width=0.5\linewidth]{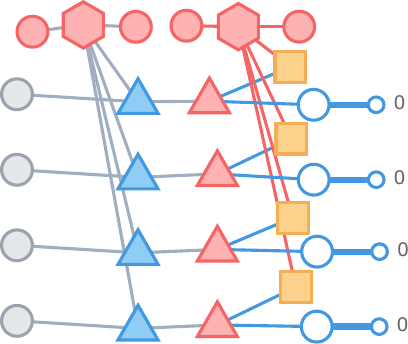}
    \caption{The MSP network for the \codepar{4,2,2} code. Logical degrees of freedom are the gray LEGOs, physical ones are the dangling legs off of the identity LEGOs (white) circles. This encoding map has a non-trivial kernel. The $ZZZZ$ stabilizer of the encoding map is highlighted through operator pushing and matching. }
    \label{fig:422-msp}
\end{figure}

The MSP circuit for a stabilizer code is a standard circuit, where physical qubits are in the $\ket{0}$ state, then stabilizer generators are measured onto an ancillary qubit, and the code space is then recovered using decoding. In the QL formalism, the random measurement outcomes are not captured, and instead, we interpret the network as a projection to the shared +1 eigenstate of the stabilizer generators. In \cref{fig:422-msp}, we display the example network for the \codepar{4,2,2} code with stabilizers $\langle XXXX,ZZZZ\rangle$. For each multi-qubit Pauli operator eigenvalue measurement, the ancilla qubit (on the top) is reset in the $\ket{+}$ state and then measured in the $\ket{+}$ state, represented by the red $X$-stoppers. The ancilla qubit will then correspond to a phase-flip code, which we display as a red LEGO. In this encoding map, alongside the four horizontal physical qubit lines, the logical legs are represented as a subspace with gray identity stoppers, and the physical legs are represented with the dangling lines with the small white circles, hanging off of the identity LEGO for emphasis. The shapes of the $CXXXX$ and $CZZZZ$ operators follow from the $CX$ and $CZ$ operator shapes discussed in \cref{sec:background}. It is easy to see how this is a universal pattern that generalizes to all stabilizer codes, with multi-qubit Pauli operator projections. Also, by contracting the ancilla repetition codes and their $X$-stoppers into a single tensor and the tensors in the qubit lines into one tensor per physical qubit, we get a graph that is isomorphic to the Tanner graph of the code. As Tanner graphs are as diverse as codes, it is hard to bound the contraction costs tightly. In fact, we find that for some codes, the contraction cost for the MSP network or the Tanner network exceeds the brute-force cost.

\section{Stabilizer code tensors are sparse} \label{sec: sparsity}

The data structure representing tensors can have crucial complexity consequences for the final space and time requirements during a contraction. For the tensor representation, it is essential to consider the average density of the tensors. A tensor is dense if most of its elements are non-zero, while a sparse tensor has only a few non-zero elements. If the average density is below some threshold, sparse representations can save on memory footprint and potentially speed up floating-point operations. Here, we show that the stabilizer code tensor networks employed for WEP calculations are sparse on average, thereby warranting a sparse representation.

\begin{figure*}[htbp!]
    \centering
    \includegraphics[width=.8\linewidth]{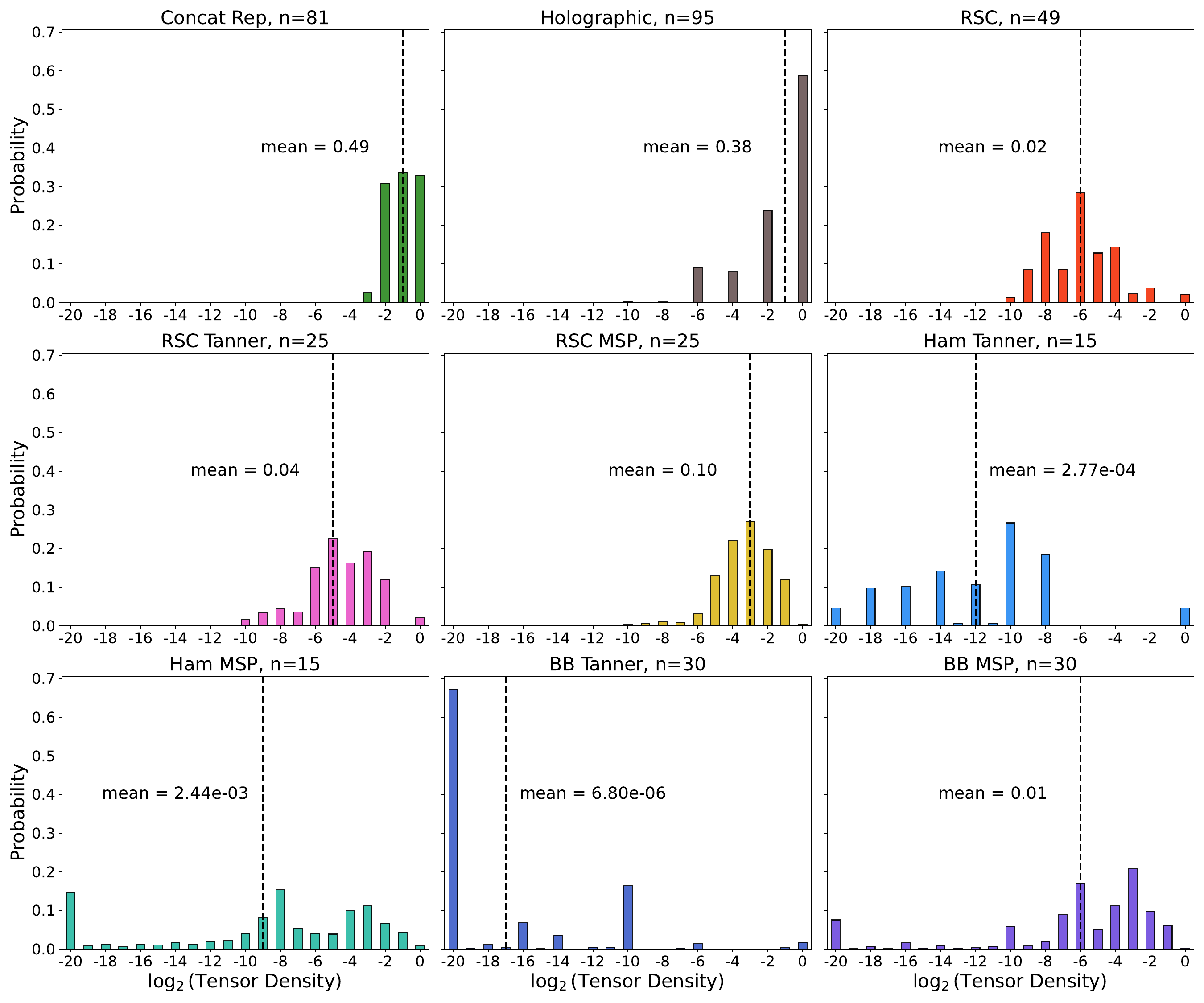}
    \caption{Density distribution of intermediate tensors during a contraction for the largest studied members of the different code families in this paper. For each code, we display the mean of the distribution.}
    \label{fig:sparsity}
\end{figure*}

To determine how sparse the tensors are for our use case, we collect the density of intermediate tensors during a tensor network contraction of different stabilizer code families. In \cref{fig:sparsity} we display the distribution of intermediate tensor density for nine codes and layouts. The average density is 0.6 for the holographic code example, 0.49 for the concatenated repetition code, but most of the other examples are significantly sparser, starting with the rotated surface code having around 0.02 average density.

The sparseness of intermediate stabilizer code WEP tensors has two consequences. Firstly, it justifies the use of sparse representations for most layouts in our sample, for example, a coordinate format (COO), or the Compressed Sparse Fiber (CSF) \cite{smithSparseTensorFactorization2017a} format. Note that the holographic and concatenated codes could potentially benefit from a dense layout. Second, the contraction cost of actual polynomial multiplications will now be only on non-zero elements, and invalidates the default Cotengra cost function's dense tensor assumption. We expect that the sparser the code/layout, the bigger the discrepancy will be.

\section{Sparse stabilizer tensor cost function for stabilizer codes} \label{sec: contraction cost function}

\begin{figure}[htbp!]
    \centering
    \includegraphics[width=1\linewidth]{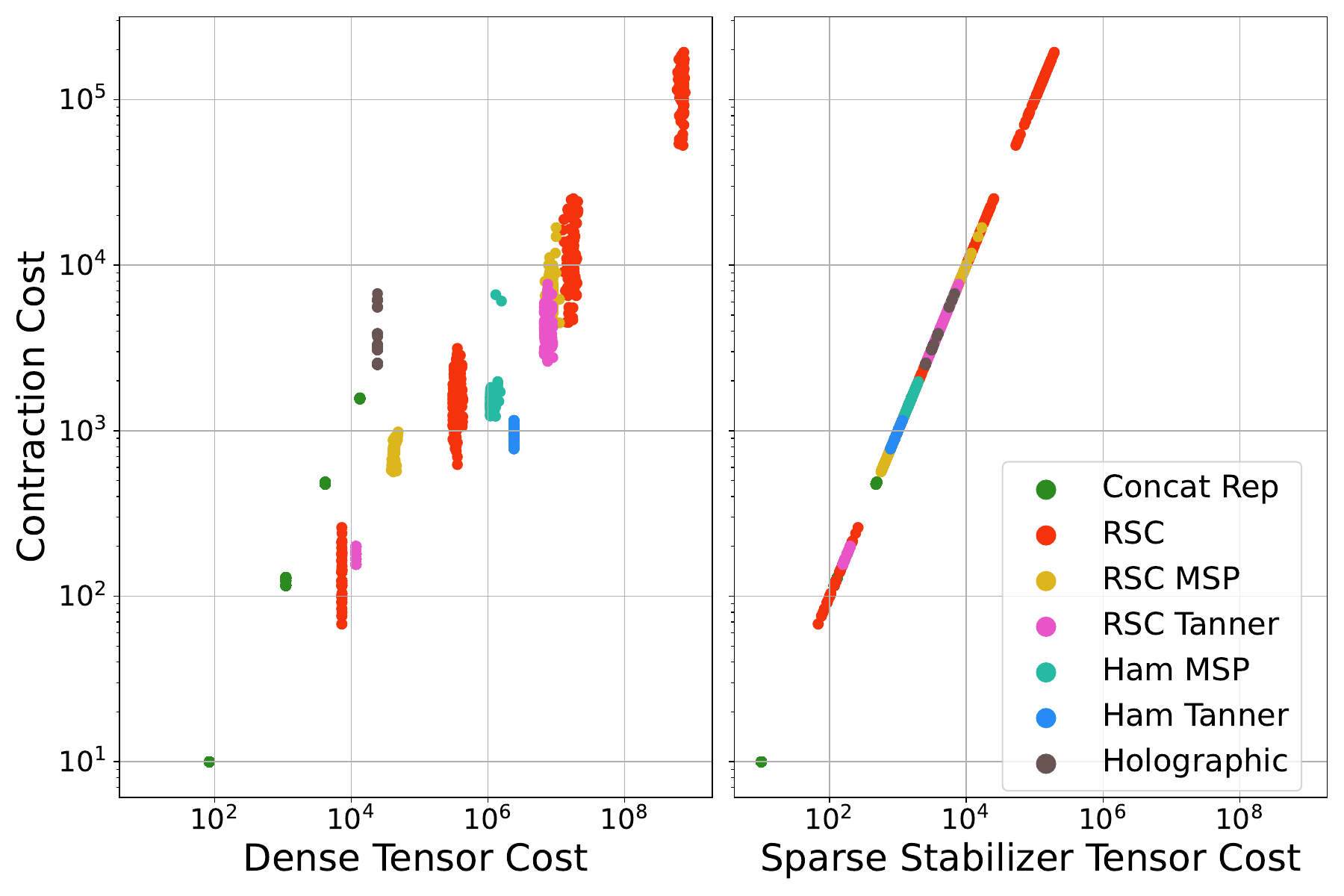}
    \caption{Contraction cost function vs true contraction cost (number of polynomial multiplications counted during contractions) for (a) Cotengra's default dense tensor cost function and (b) our sparse stabilizer tensor cost function. The dense tensor cost function, although it correlates with, fails to accurately estimate the true cost function. The core insight leveraged by our cost function is that we have a polynomial complexity algorithm to determine the number of non-zero elements in the weight enumerator polynomial tensors for stabilizer codes as described in \cref{sec: contraction cost function}. This translates to significant speedups for most cases examined in \cref{sec:results}.}
    \label{fig:compare cost functions}
\end{figure}

As demonstrated in \cref{sec: sparsity}, the tensors in our stabilizer code tensor networks stay sparse throughout the contraction. Here, we describe a polynomial-time algorithm to determine the non-zero elements in the intermediate tensors, as well as an algorithm to determine the number of matching elements during contraction. The backbone of our algorithm is the parity check matrix calculation algorithm described in \cref{sssec: QL PCM}.

For a given contraction tree, we can estimate the contraction cost by executing the contraction and keeping track of the PCM of intermediate tensors. For a WEP calculation, as described in \cref{sssec: QL WEP}, at every step we contract two separate tensors on some legs. The number of operations in a contraction of two tensors is taken to be the number of polynomial multiplications. Our sparse implementation of the algorithm for WEP calculation described in \cref{sssec: QL WEP} for a given contraction tree is simply iterating through all the possible basis elements $e_{E_1 \otimes E_2}$ in the tensor product, but only calculating the multiplication $A_{E_1}^{(J_1)}(z;\Pi_{C_1}) A_{E_2}^{(J_2)}(z;\Pi_{C_2})$ between the two polynomials for \textit{matching elements}. A matching element has the same operators on the corresponding indices defined by the contraction, or formally, for a contraction between the indices $A=\{(a_1, a_2) | a_1 \in J_1, a_2 \in J_2 \}$, a matching element has the property that for all $(a_1, a_2) \in A: (E_1)_{a_1} = (E_2)_{a_2}$. We count these elements using an $O(n^3)$ algorithm below.

Firstly, we know that $rank(H)$ of the PCM $H$ tells us exactly the number of independent generators that generate the stabilizer group for that given code. As we have an Abelian group, the order of the stabilizer group will be $|S|=2^{r}$. Similarly, we can determine the number of elements in a tensor WEP with open indices $J, |J|=k$, based on the rank of $H|_J$, the submatrix of the PCM (sub-PCM) on these legs, as $2^{rank(H|_J)}$.

Second, we'll determine the $R_m$ ratio of elements in the tensor product of two PCMs that match on a given set of pairs of indices. Let the two sub-PCMs we are joining be $H_1|_{J_1}$ and $H_2|_{J_2}$, with $|J_1| = |J_2|$, the lists of indices we are joining the two tensors on. Matching elements are the rows in $B \equiv \left( \begin{array}{c|c}
            H_1|_{J_1} & 0          \\
            0          & H_2|_{J_2}
        \end{array}\right)
$ that have the same operator on the first half as in the other half. Clearly, this is exactly the intersection of the rowspace of $H_1|_{J_1}$ and the rowspace of $H_2|_{J_2}$, which we'll denote with $I \equiv row(H_1|_{J_1}) \cap row(H_2|_{J_2})$. Then,

\begin{align*}
    |row(B)| & = 2^{rank(H_1|_{J_1})}2^{rank(H_2|_{J_2})}
\end{align*}

as the generators of the rowspace of $B$ are disjoint on the two halves of $H_1|J_1$ and $H_2|J_2$. In contrast, we define the matrix

$$
    W \equiv \left(\begin{array}{c}
            H_1|_{J_1} \\
            H_2|_{J_2}
        \end{array}\right),
$$

\noindent the rows of which will generate the union of elements generated by the two sub-PCMs. To count these, we have to remove the intersection to avoid double-counting, and hence:

\begin{align*}
    |row(W)| & = \frac{2^{rank(H_1|_{J_1})}2^{rank(H_2|_{J_2})}}{|I|}.
\end{align*}

But what we are looking for is exactly the ratio

$$R_m=\frac{|I|}{2^{rank(H_1|_{J_1})}2^{rank(H_2|_{J_2})}},$$

\noindent and thus, we can calculate $2^{-rank(W)} = R_m$ as our ratio.

Finally, for the total number of polynomial multiplications, we get $C=2^{rank(H_1|_{J_1})+rank(H_2|_{J_2})}R_m$. The total cost to calculate is $O(n^3)$, as calculating the rank requires Gaussian elimination. We compare the time cost of the Sparse Stabilizer Tensor (SST) to Cotengra's dense tensor cost function in \cref{app: cost function timings}.

To examine how well the SST cost function correlates with the actual cost, we contrast it with Cotengra's default dense tensor cost function, and plot the estimated cost vs the actual contraction cost in \cref{fig:compare cost functions}. The dense tensor cost (called the \texttt{flops} cost function in Cotengra) demonstrates a considerable amount of uncertainty, assigning the same cost to a wide range of contraction trees. In contrast, the SST cost function correlates perfectly with the actual contraction cost over the example QL networks.

\section{Performance results} \label{sec:results}

We investigate the impact of using SST versus the dense tensor cost function on actual contraction costs after the Cotengra contraction optimization across 20 codes. We explore 6 MSP and Tanner networks: the Quantum Hamming codes \href{https://errorcorrectionzoo.org/c/quantum_hamming_css}{\eczooUseIcon} \cite{steaneSimpleQuantumError1996} \codepar{7,1,3} and \codepar{15,7,3}, the rotated surface code (RSC) \href{https://errorcorrectionzoo.org/c/rotated_surface}{\eczooUseIcon} \cite{bombinOptimalResourcesTopological2007} for $n=9$ and $n=25$, and two members for the bivariate bicycle (BB) codes \href{https://errorcorrectionzoo.org/c/qcga}{\eczooUseIcon} of \codepar{18,4,4} \cite{wangDemonstrationLowoverheadQuantum2025} and \codepar{30,4,5} \cite{yeQuantumErrorCorrection2025a}. Then, for the two-dimensional grid network, we look at the RSC of distances 3, 5 and 7 with $n=9$, $n=25$ and $n=49$, respectively, based on \cite{caoQuantumLegoExpansion2024a}. For holographic codes, we use $n=25$ and $n=95$ members of the HaPPy code family \href{https://errorcorrectionzoo.org/c/happy}{\eczooUseIcon}\cite{pastawskiHolographicQuantumErrorcorrecting2015a}. And finally, we use the family of concatenated repetition codes with $n=9$ (Shor's code), $n=27$, and $n=81$.

\begin{figure*}[!htbp]
    \centering
    \includegraphics[width=1\linewidth]{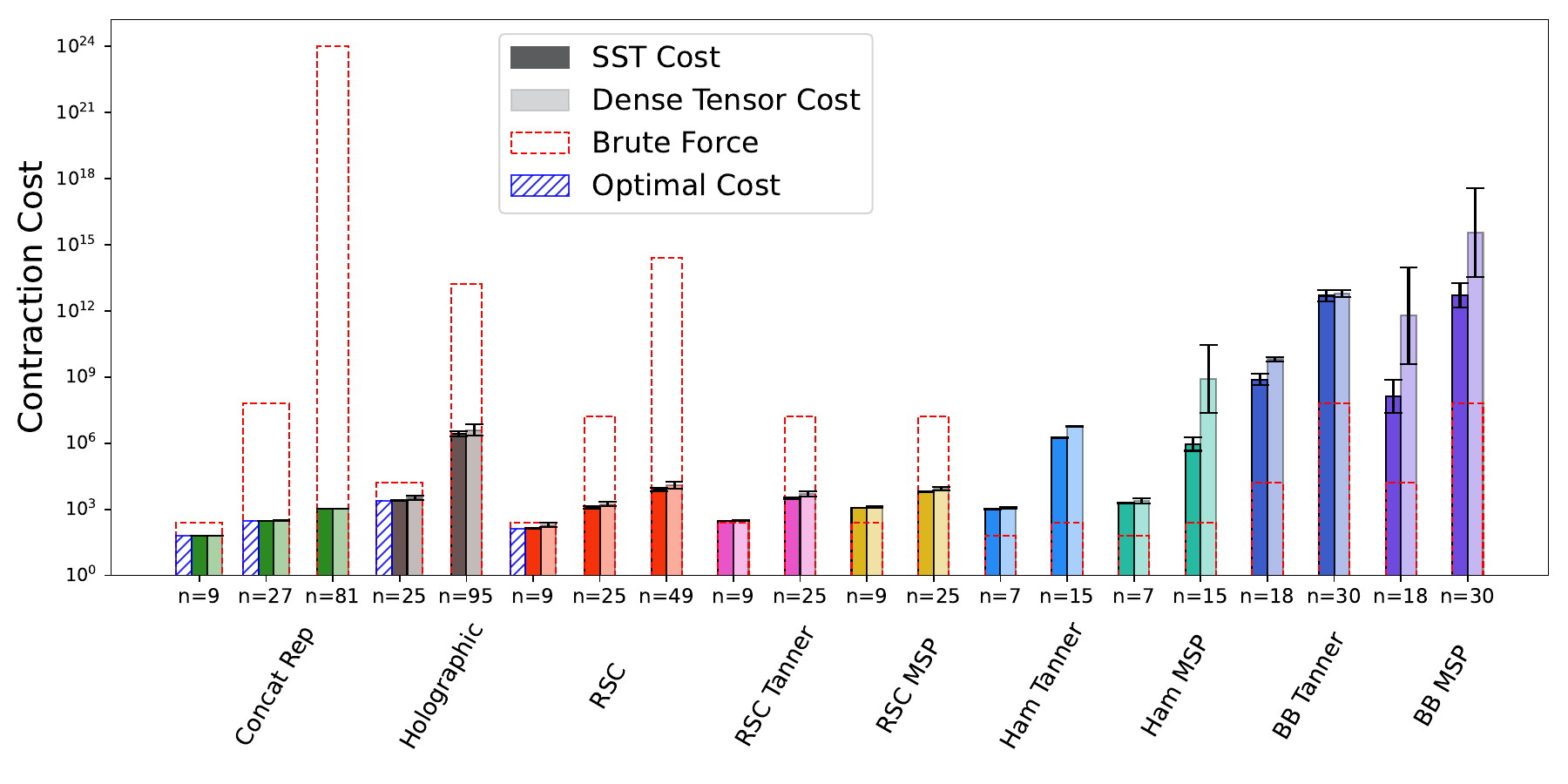}
    \caption{The \texttt{Hyper-Greedy} algorithm on QL networks using the default Cotengra cost function and the sparse stabilizer tensor cost function. The red dotted bars indicate the cost required to calculate the WEP using brute-force counting. The number above the bars indicates the ratio of dense tensor cost operations to SST operations, showing the speedup factor. For small examples, we compared the results to the output of the slow \texttt{Optimal} strategy (dashed). It is clear that Cotengra reaches close to optimal contraction trees in these cases. The bar heights are the geometric mean of the contraction costs, and the error bars are the geometric standard deviation. We choose to use the geometric mean given the skew in the data.}
    \label{fig:greedy}
\end{figure*}

For each of the 20 codes, we perform 100 runs of Cotengra's random algorithm and record the mean and the standard deviation. We run both the default dense flops cost function and the SST cost function, using \texttt{Hyper-Greedy}, \texttt{Hyper-Par} (with KaHyPar \cite{KaHyPar1,KaHyPar2}), and \texttt{Optimal} hyper-optimization strategies. \texttt{Hyper-Par} is the slower of the two strategies, and thus, it was not feasible to calculate the contraction schedule for the $n=30$ BB code. We note that for the deeper levels of \texttt{Hyper-Par}, we also use the SST cost function. For \texttt{Hyper-Greedy}, the size calculation is still using a dense assumption, and as such, we expect more improvement by implementing a rank-based calculation. See details in \cref{app: greedy tensor size}.

As the SST cost function is slower, we compare them using two different cutoff strategies: limiting the Cotengra trials to 64 or limiting the execution time to 5 minutes. Here, we explain the results of the 64 trials. Since the 5-minute cutoff pertains to timing differences, those results are shown in \cref{app: cost function timings}.

\begin{figure*}
    \centering
    \includegraphics[width=1\linewidth]{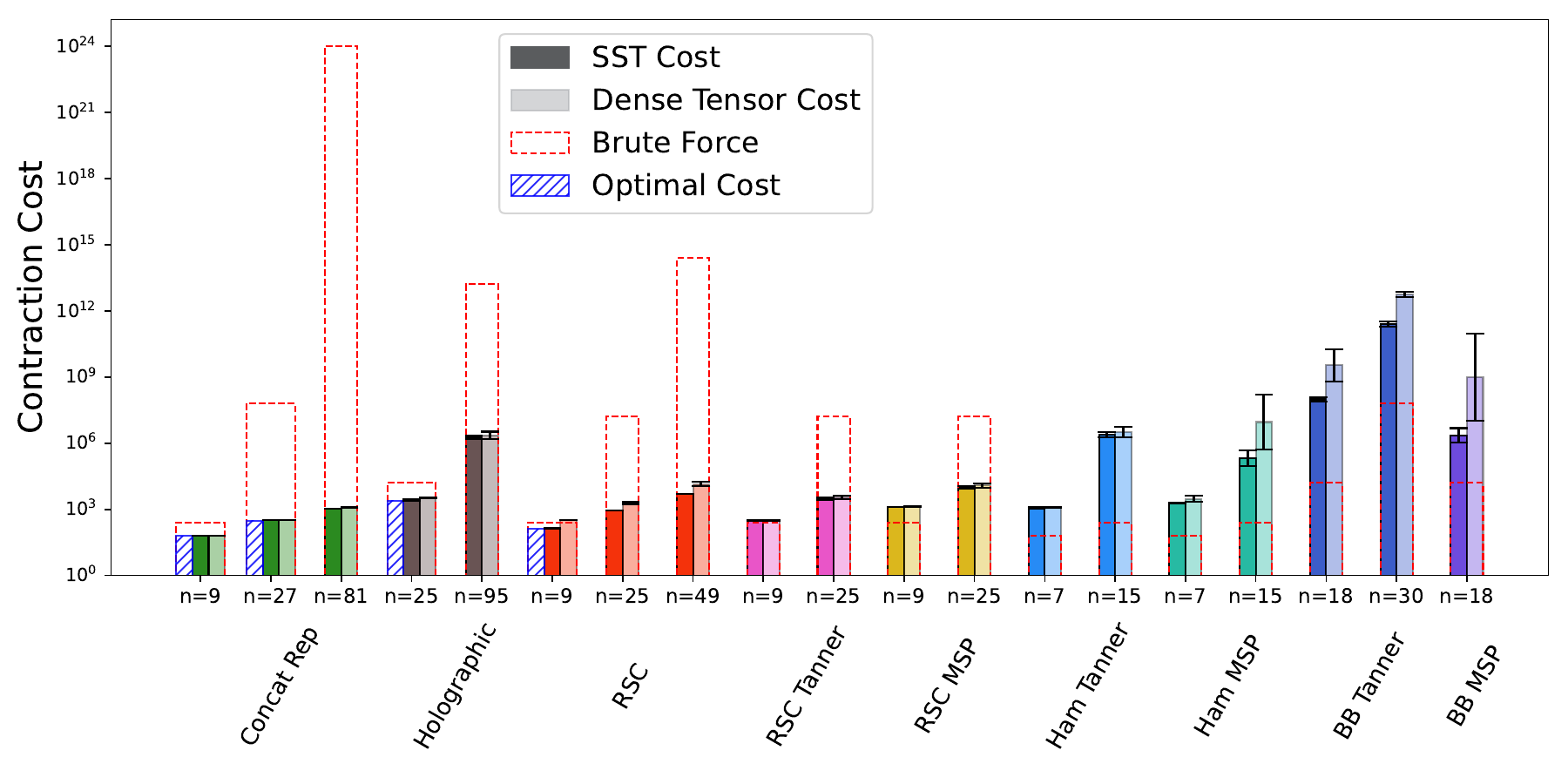}
    \caption{The \texttt{Hyper-Par} algorithm on QL networks using the default Cotengra cost function and the sparse stabilizer tensor cost function. The red dotted bars indicate the cost required to calculate the WEP using brute-force counting. The number above the bars indicates the ratio of dense tensor cost operations to SST operations, showing the speedup factor. For small examples, we compared the results to the output of the slow \texttt{Optimal} strategy (dashed). It is clear that Cotengra reaches close to optimal contraction trees in these cases. The bar heights are the geometric mean of the contraction costs, and the error bars are the geometric standard deviation. We choose to use the geometric mean given the skew in the data.}
    \label{fig:kahypar}
\end{figure*}

The performance results for \texttt{Hyper-Greedy} are shown in Fig. \ref{fig:greedy} and for \texttt{Hyper-Par} in Fig. \ref{fig:kahypar}. The average values and improvement factors are shown in Appendix \ref{app: cost function table}.

The SST cost function generally finds contraction schedules with lower contraction costs than the default cost function, except for the $n=9$ Concatenated Repetition code, for which the cost functions demonstrate equal performance. For a small tree tensor network with relatively high average tensor density, it is not surprising that both cost functions can find a solution very close to optimal.

For most of the layouts using a universal representation (MSP and Tanner), tensor network contraction performs worse than directly calculating the WEP through brute force. The exceptions are the RSC MSP with $n=25$ and RSC Tanner with $n=25$ layouts, where both the SST cost function and the dense tensor cost function reduce the contraction cost to below that of brute force. For all other MSP and Tanner representations, both the SST and default cost functions perform worse than brute force, and therefore, even though SST would improve the contraction cost significantly, using brute force would be faster.

We also examined the variability of the results by comparing the standard deviation found across all runs. In most cases, the SST cost function reduced variability, indicating more stable performance. One exception is the BB Tanner code, where the SST does achieve a lower average cost, but the variability is larger than the default.

The \texttt{Hyper-Par} method outperforms \texttt{Hyper-Greedy} for BB MSP, BB Tanner, Hamming MSP, Holographic ($n=95$), and RSC ($n=25$ and $n=49$). \texttt{Hyper-Greedy} wins in all other cases. The relative improvement factors of \texttt{Hyper-Par} and \texttt{Hyper-Greedy} vary across codes, with neither method consistently achieving the greater speedup.

Surprisingly, despite consisting of significantly more tensors, the MSP networks outperform the Tanner networks in the cases of $n=15$ quantum Hamming code and the $n=18$ bivariate bicycle codes.

All source code and data are available for download at \cite{vanlerberghe_2025_17290496}.

\section{Conclusions and outlook}\label{sec:conclusions}

Quantum LEGO generalizes code concatenation and provides an alternative approach to exploring the design space of QEC codes. It also features weight enumerator polynomial calculation via tensor network contraction. The complexity of calculating the WEP via quantum LEGO depends on the entanglement structure of the code, the layout choice, and the contraction schedule. We find that the framework for hyper-optimized contraction schedules works well with the defaults provided by the Cotengra library, which we can then optimize further using our SST cost function for stabilizer codes. Our cost function also reduces uncertainty in the convergence of the Cotengra strategies and provides an exact number of operations for our sparse tensor networks. We find a correlation between the sparsity of the intermediate tensors in the network and the speedup factor of our cost function.

The SST cost function is efficient, scales as $O(n^3)$, and is exact, and thus can help decide whether calculating the WEPs for a given code and QL layout is better via brute force methods or using a given contraction schedule. Most of our Tanner and MSP layout examples for our 64-trial cutoff are beyond the cost of brute force, except for the case of the larger, $n=25$ rotated surface code instances. It is possible that increasing the sample size of the Cotengra hyperoptimizer would decrease the cost of contraction. The MSP layout, in particular, features numerous small tensors, resulting in a larger space for the hyperoptimizer to explore. Consequently, more trials could improve the quality of the contraction trees. It is unclear whether an optimal layout would be able to improve the cost below brute force. The entanglement structure of the code may render them non-viable for QL contraction. Namely, volume-law entanglement in these codes would cause beyond-brute force contraction costs \cite{caoQuantumLegoExpansion2024a}. There is recent evidence for the BB codes having worse-than-area-law entanglement \cite{zhaoGraphbasedApproachEntanglement2025a}, which could explain the worse-than-brute force contraction costs. Furthermore, the area law scaling of the rotated surface code might be the very reason for the RSC contraction cost's better-than-brute force scaling for these universal layouts. The analysis and quantification of contributing factors are of great interest and will be the subject of follow-up work.

Our cost function assumes the naive contraction algorithm, where all the traces between tensors are done sequentially, even if there are repeated subproblems. Hence, tailor-made algorithms that can exploit the repeated substructures in problems are expected to perform better. Expanding our cost function to account for these types of optimizations is left for future work.

Approximate contraction for tensor networks holds the promise for reducing computational cost if the target computation can tolerate a loss of accuracy up to a certain level of tolerance. It is of interest to explore these contraction schedules for decoding, for example, where approximate contraction is effective \cite{bravyiEfficientAlgorithmsMaximum2014a}. For some tensor networks, the decoding use case might allow for other optimizations, including parallelization for multiple qubits \cite{steinbergUniversalFaulttolerantLogic2025, farrellyParallelDecodingMultiple2022}. Exploring hyper-optimized contraction schedules for these use cases would be valuable.

For the more general weight enumerators, as described by Cao and Lackey \cite{caoQuantumWeightEnumerators2024}, such as the complete and general tensor weight enumerators, we expect that our methods will translate well; an explicit analysis is of future interest.

The SST cost function is a promising tool for probing further pressing open questions that will increase the usability of the quantum LEGO framework for code design, including finding optimal QL layouts. The surprising result, where the MSP network outperforms the Tanner network for larger code members, supports this direction. We conjecture that for even larger members, the trend would hold, but more experiments are needed to verify and understand the effect in general. It is possible that, as the Tanner network can be thought of as a coarse-grained version of the MSP network, it narrows down the possible contraction paths. The ones that are left are more expensive on average than the ones Cotengra finds starting from the MSP network.

The SST cost function, alongside Cotengra's hyper-optimized contraction schedules, increases the potential for QL to find new classes of codes for which the entanglement structure and optimal QL layouts are viable for WEP calculation. Our tools thus aid in code design, and we are hopeful that they will lead to the expansion of the class of practical, WEP-viable codes.

\begin{acknowledgments}
    We acknowledge useful discussions with Charles Cao and Brad Lackey. The authors acknowledge support from the Unitary Fund, the Office of
    the Director of National Intelligence (ODNI), Intelligence
    Advanced Research Projects Activity (IARPA), under
    the Entangled Logical Qubits program through Cooperative Agreement Number W911NF-23-2-0216 and the
    ARO/LPS Quantum Characterization of Intermediate Scale Systems program through Grant Number W911NF-21-1-0005.
\end{acknowledgments}


\bibliographystyle{quantum}
\bibliography{refs.bib}

\newpage
\clearpage

\appendix

\section{Sparsity metrics for smaller codes}

\begin{figure*}[htbp!]
    \centering
    \includegraphics[width=1\linewidth]{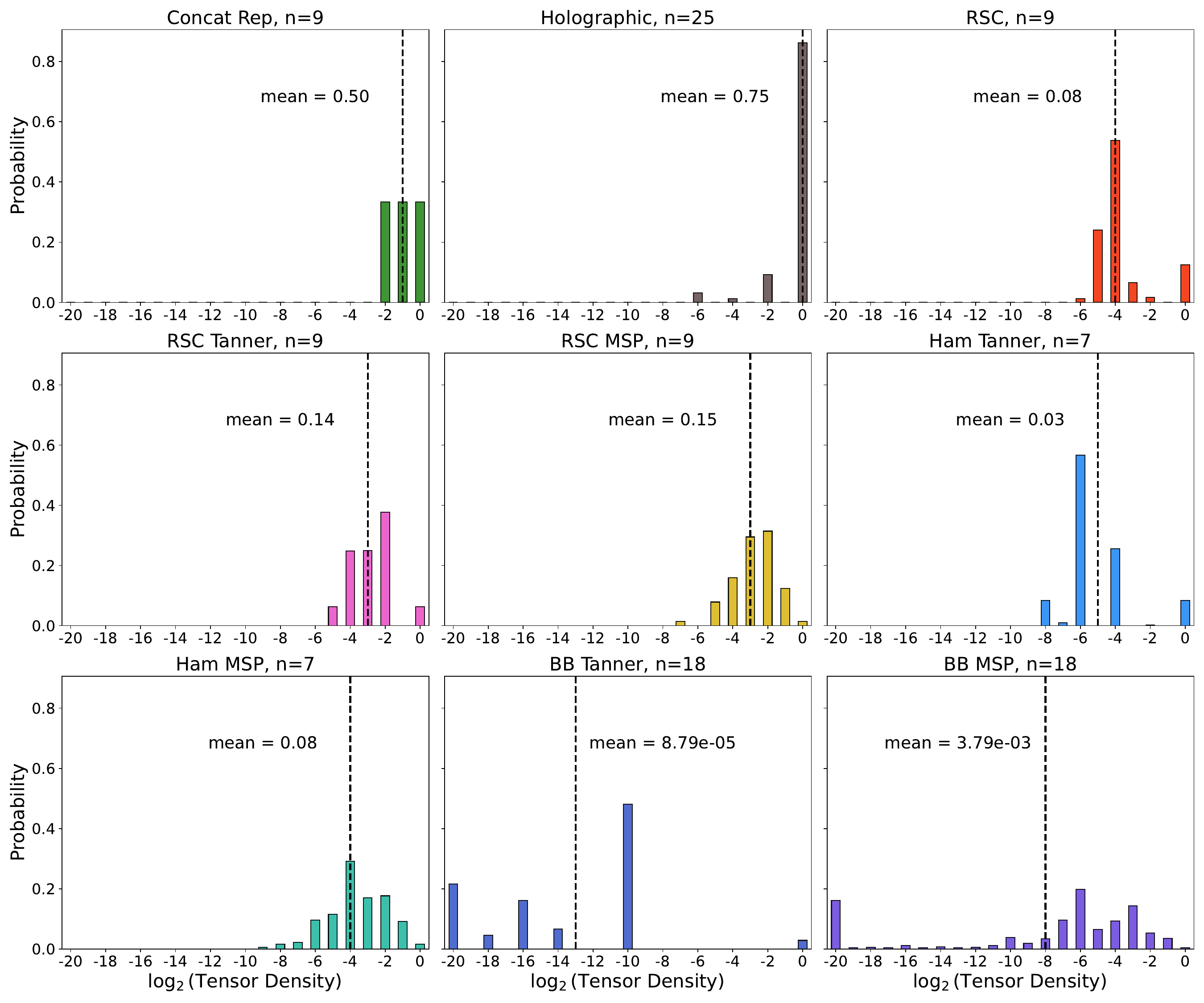}
    \caption{Sparsity of intermediate tensors during tensor network contraction for small codes. The mean is shown on the plot and represented by the dotted vertical line. }
    \label{fig:small code density}
\end{figure*}

As compared to \cref{fig:sparsity}, we see that in \cref{fig:small code density} the smaller codes are typically less sparse than the larger ones.

\section{Computing the tensor size in the \texttt{Hyper-Greedy} strategy} \label{app: greedy tensor size}

\begin{figure*}[!htbp]
    \centering
    \includegraphics[width=0.8\linewidth]{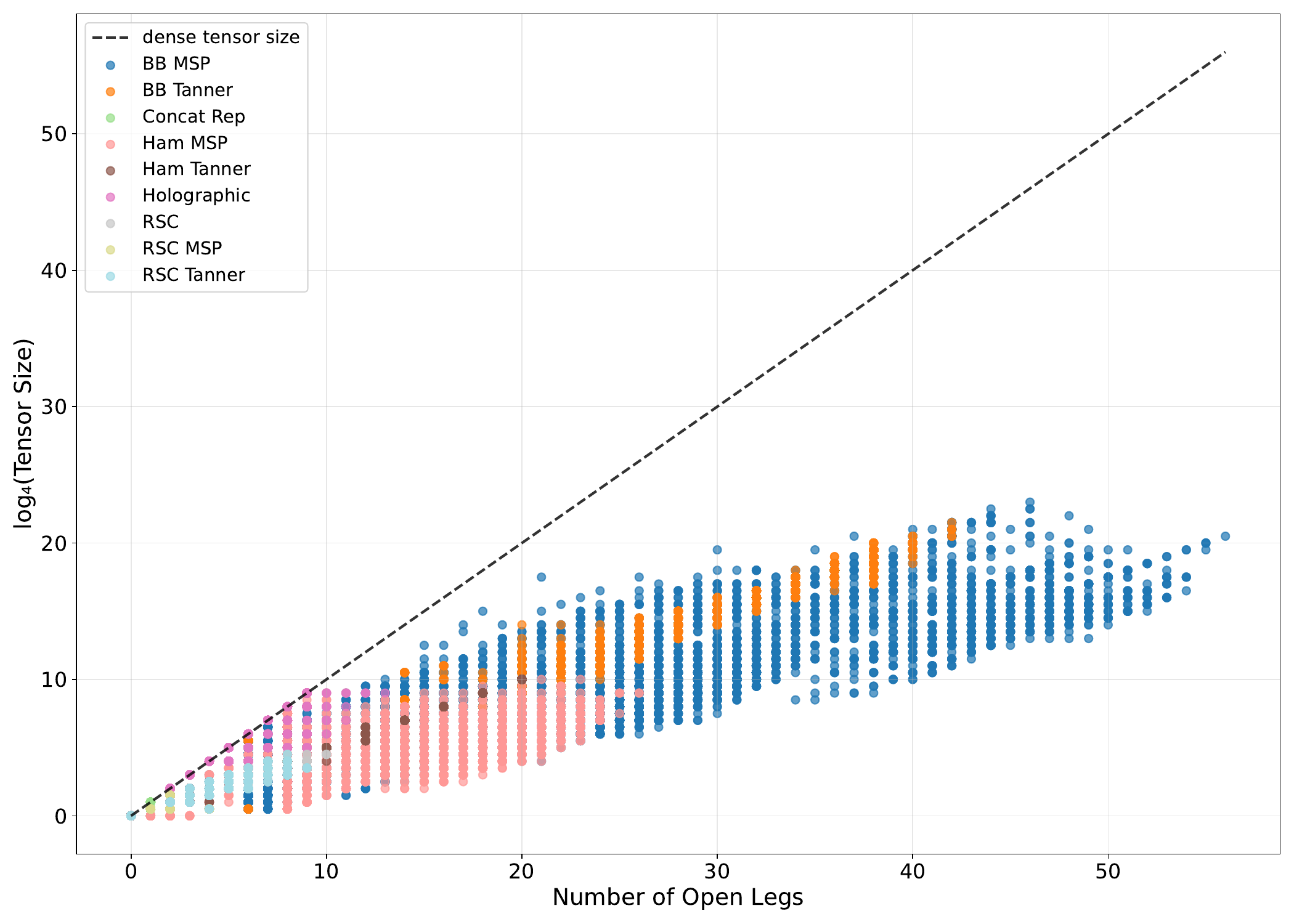}
    \caption{The $log_4$ of the true tensor size (using rank) vs the number of open legs for the different studied families in the main text. We can see that a lot of the intermediate tensors fall below the dense tensor line, and in many cases, choosing a tensor that has fewer open legs doesn't imply smaller size (in contrast with the strictly monotonic dense tensor metric). }
    \label{fig: sparsity line}
\end{figure*}

While the \texttt{Hyper-Greedy} strategy in our implementation uses the SST cost function for the hyper-optimizer part, it internally uses the dense tensor assumption for determining the tensor sizes \cref{eq: greedy equation}. This could be further improved by using the rank instead, as it is used in the SST cost function described in \cref{sec: contraction cost function}. In \cref{fig: sparsity line}, we display the log of the true size of intermediate tensors versus the number of legs. This shows that there are plenty of cases where, using the dense assumption, the Greedy function would pick tensors with less open legs, but that might actually have a higher actual size based on the rank.

However, adding this slowed down the numerics significantly and, having good results without this improvement, we left for future work to implement a more efficient rank calculation combined with DP methods alongside the greedy algorithm.

\section{Cost function time analysis} \label{app: cost function timings}

\begin{figure*}
    \centering
    \includegraphics[width=1\linewidth]{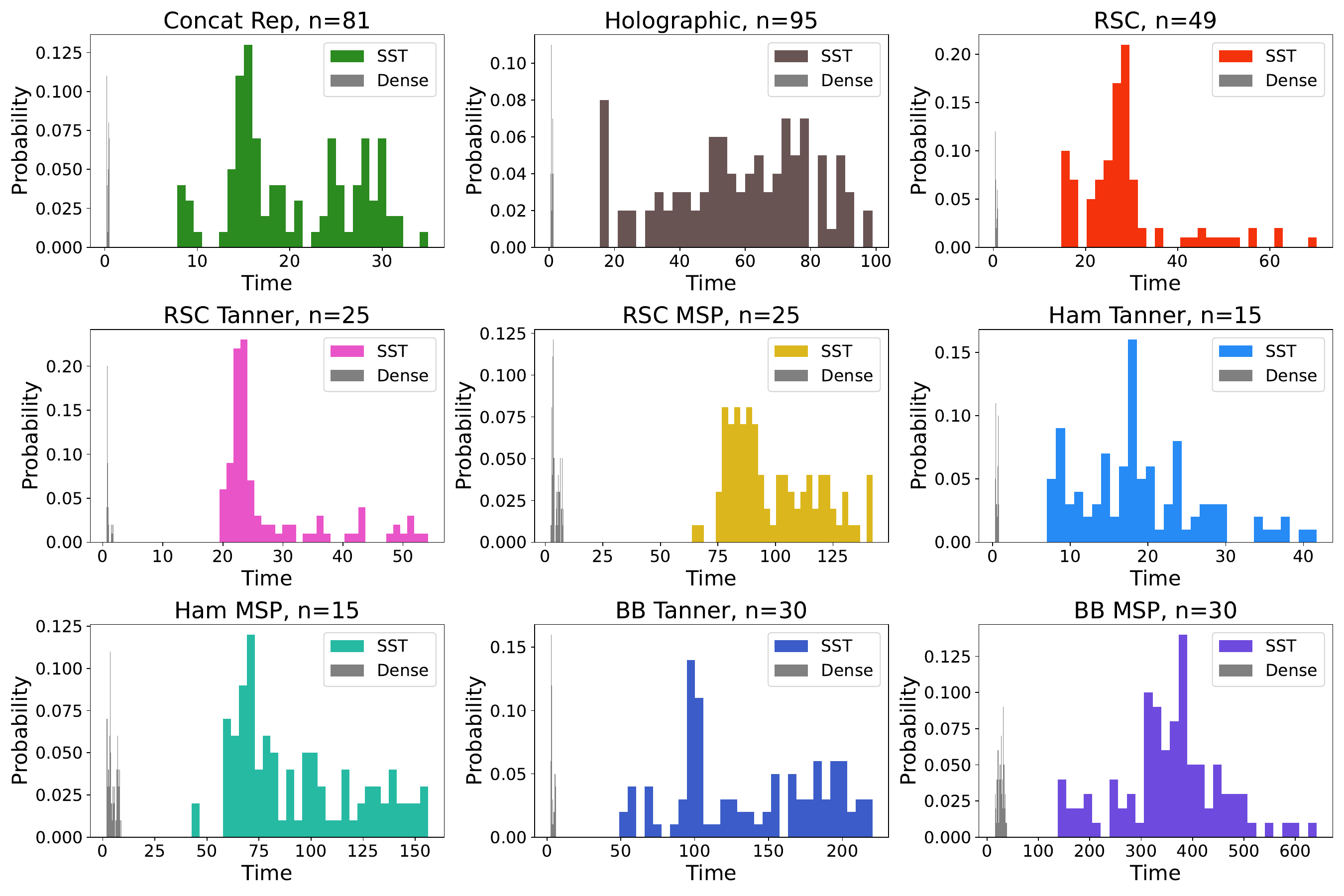}
    \caption{Comparison of the computation time for the SST cost function vs the default dense cost function for the \texttt{Hyper-Greedy} method. }
    \label{fig:time comparison}
\end{figure*}

\begin{figure*}
    \centering
    \includegraphics[width=1\linewidth]{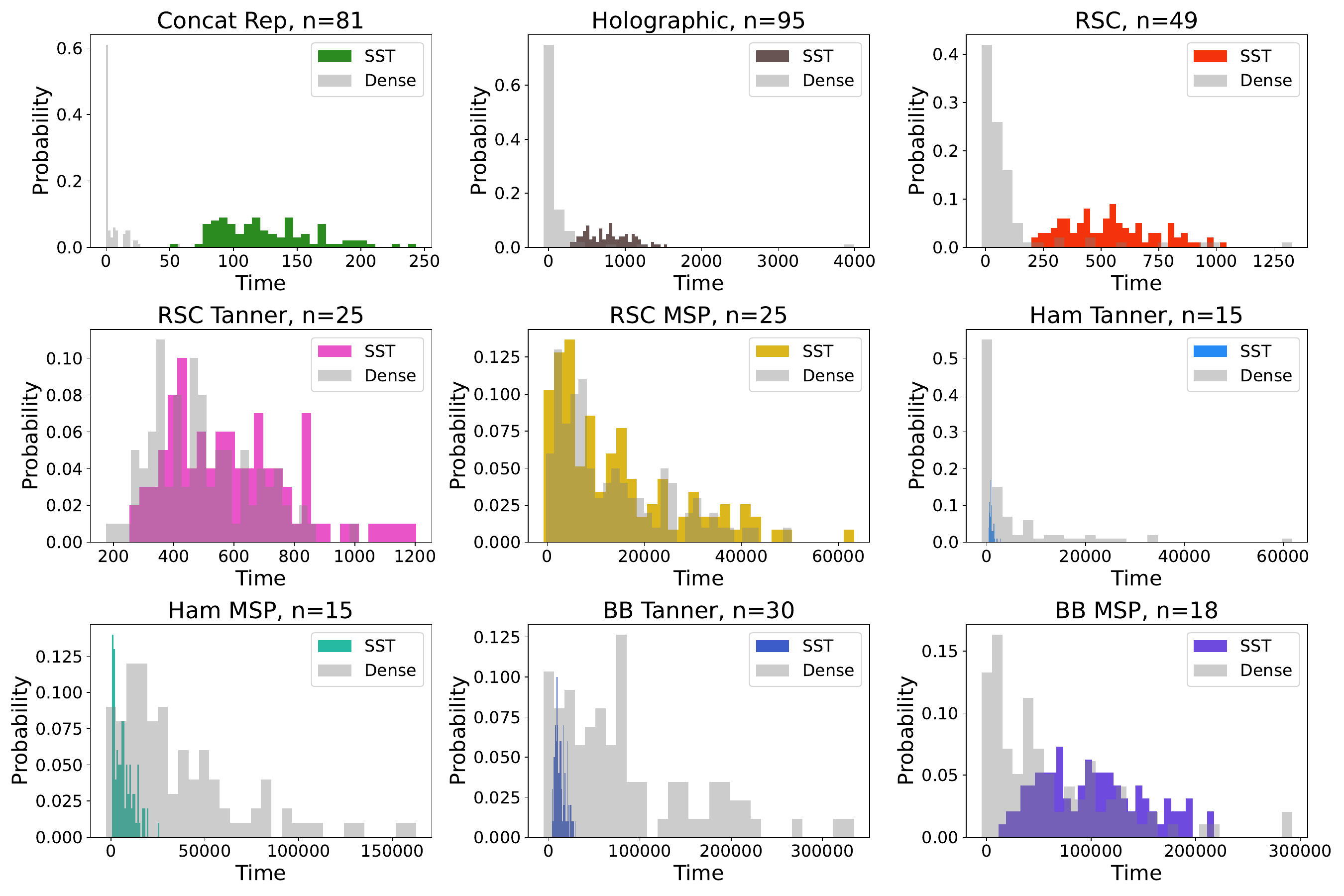}
    \caption{Comparison of the computation time for the SST cost function vs the default dense cost function for the \texttt{Hyper-Par} method. }
    \label{fig:time comparison hyperpar}
\end{figure*}

\begin{figure*}
    \centering
    \includegraphics[width=1\linewidth]{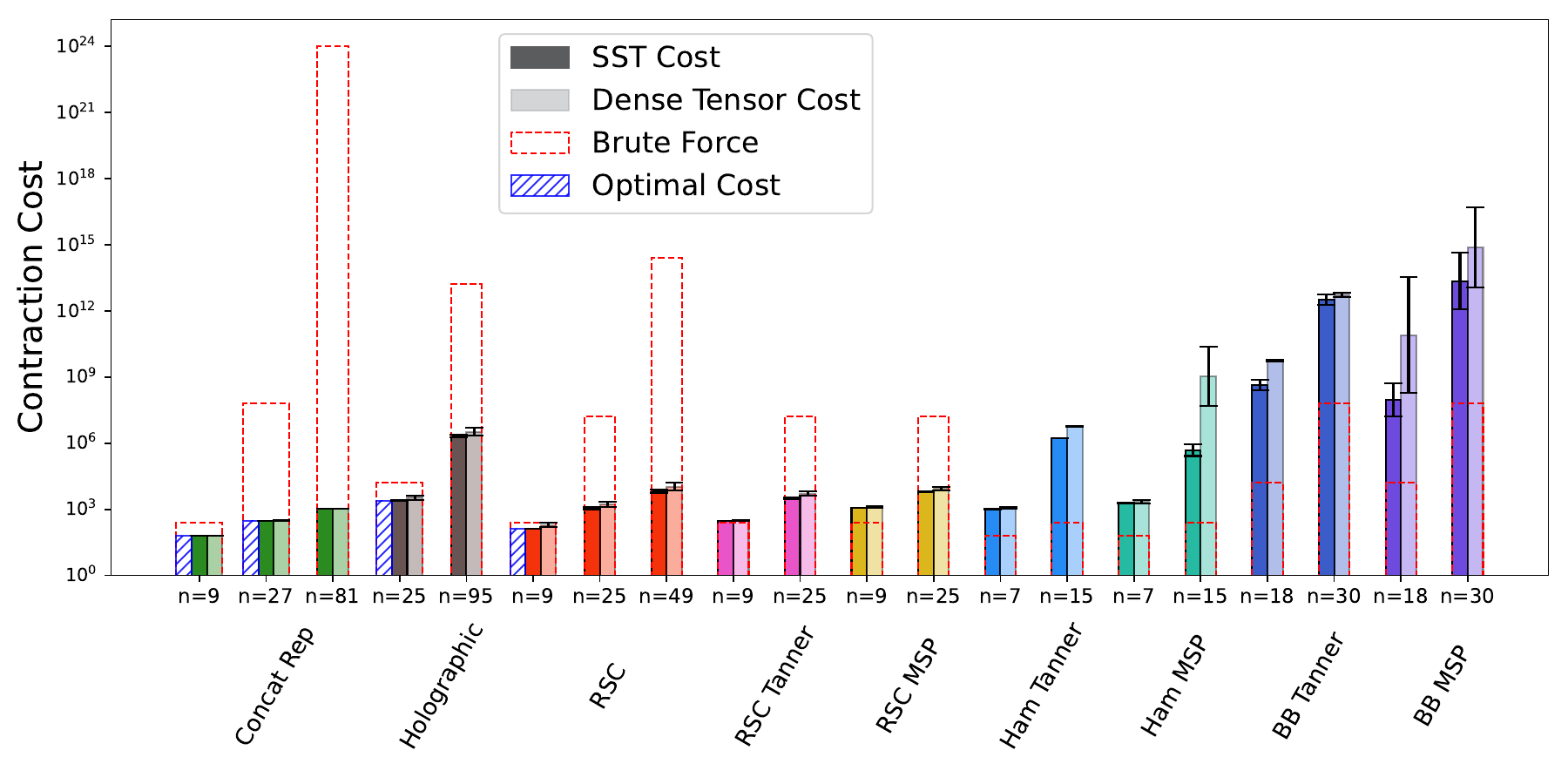}
    \caption{The \texttt{Hyper-Greedy} algorithm on QL networks using the default dense cost function and the SST cost function with a 5-minute cutoff. The red dotted bars indicate the cost required to calculate the WEP using brute-force counting. The number above the bars indicates the ratio of dense tensor cost operations to SST operations, showing the speedup factor. For small examples, we compared the results to the output of the slow \texttt{Optimal} strategy (dashed). It is clear that Cotengra reaches close to optimal contraction trees in these cases. The bar heights are the geometric mean of the contraction costs, and the error bars are the geometric standard deviation. We choose to use the geometric mean given the skew in the data.}
    \label{fig:5min bar chart greedy}
\end{figure*}

\begin{figure*}
    \centering
    \includegraphics[width=1\linewidth]{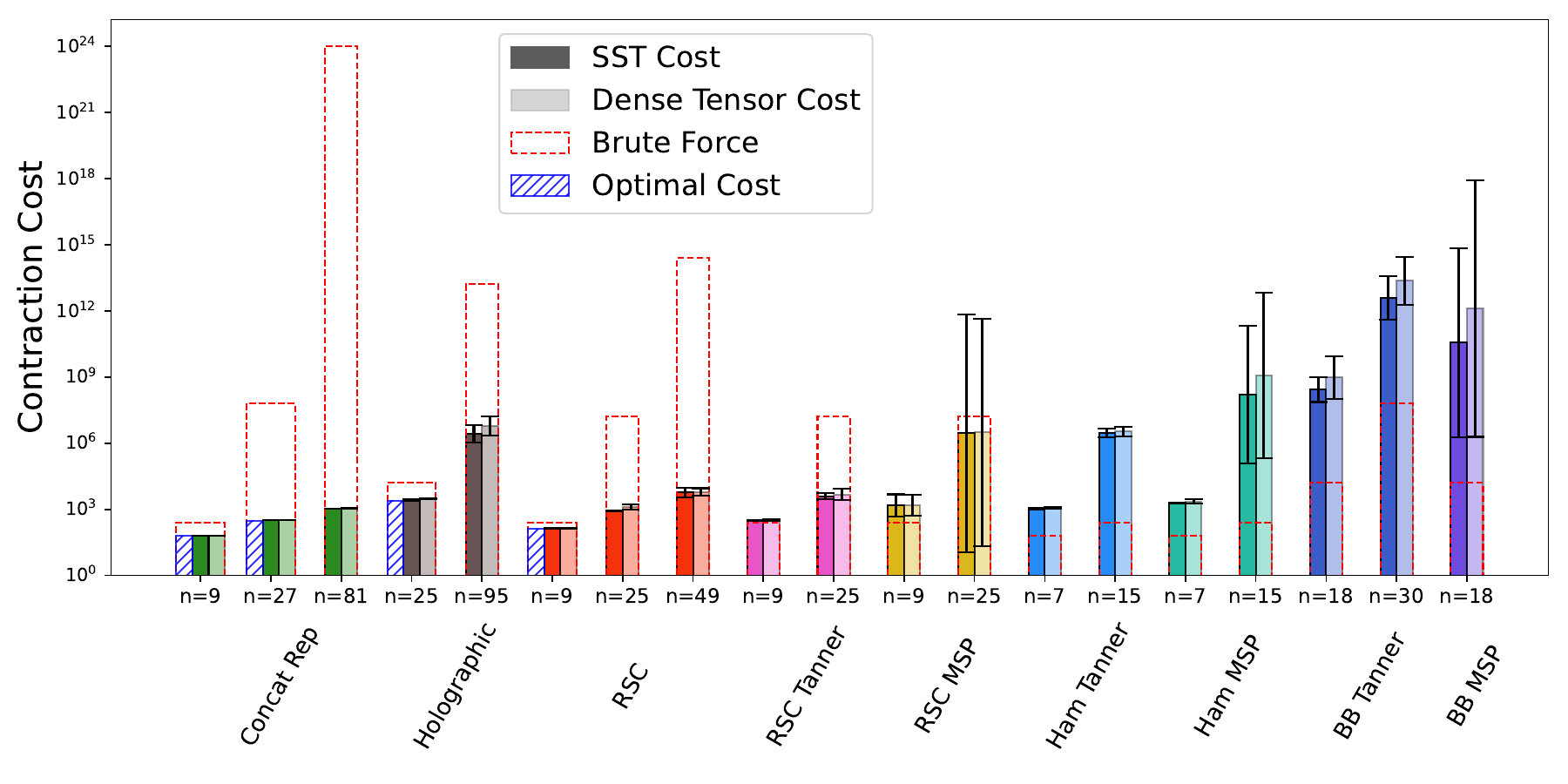}
    \caption{The \texttt{Hyper-Par} algorithm on QL networks using the default dense cost function and the SST cost function with a 5-minute cutoff. The red dotted bars indicate the cost required to calculate the WEP using brute-force counting. The number above the bars indicates the ratio of dense tensor cost operations to SST operations, showing the speedup factor. For small examples, we compared the results to the output of the slow \texttt{Optimal} strategy (dashed). It is clear that Cotengra reaches close to optimal contraction trees in these cases. The bar heights are the geometric mean of the contraction costs, and the error bars are the geometric standard deviation. We choose to use the geometric mean given the skew in the data.}
    \label{fig:5min bar chart kahypar}
\end{figure*}

Computation times for the SST and default dense cost functions are shown in \cref{fig:time comparison} for the \texttt{Hyper-Greedy} method and in \cref{fig:time comparison hyperpar} for Hyper-Par.

For \texttt{Hyper-Greedy}, we see that the use of the SST cost function increases computation time by at most 10 minutes across all codes, and is typically only about 1-3 minutes slower. For the smaller codes where the overall contraction cost is already low and thus the actual contraction time is a matter of seconds, the additional time from using the SST cost function may not be justified. However, for larger codes, this slight overhead can lead to large savings for the actual contraction. For example, in the $n=95$ Holographic code, using the default dense cost function results in an average total contraction time of 24.48 minutes (where the total contraction time includes time from the cost function to find the schedule plus the actual contraction). When using the SST cost function, even though it is slightly longer for the contraction tree optimization, the total time is only 4 minutes on average.

For \texttt{Hyper-Par}, we see a larger range of computation times for both the SST and default dense cost functions. This is because the \texttt{Hyper-Par} strategy is a recursive ensemble of optimizers, and every level adds an extra level of uncertainty. The additional $O(n^3)$ overhead of the SST exacerbates the effect.

To evaluate the performance of the SST and default dense cost functions in terms of time, we ran 100 runs of each code with a 5-minute cutoff instead of a fixed number of trials. The results are shown in \cref{fig:5min bar chart greedy} for \texttt{Hyper-Greedy} and in \cref{fig:5min bar chart kahypar} for \texttt{Hyper-Par}. Since  \texttt{Hyper-Greedy} still runs quickly, we observe similar improvements to those seen in the experiment with a fixed number of trials. The SST cost function finds lower contraction costs across all codes, except for the one equal case, as expected. For \texttt{Hyper-Par}, we observe that the default dense cost function performs similarly to the SST cost function. This is because the SST cost function does not run as many trials as the default dense cost function within the 5-minute cutoff limit.

\section{Performance Metrics for SST and Dense Cost Functions} \label{app: cost function table}

The average contraction costs found for the two cost functions are shown in Table \ref{table: costs for greedy} for \texttt{Hyper-Greedy} and in Table \ref{table:costs for hyperpar} for \texttt{Hyper-Par}. The improvement factor is also shown.

\begin{table*}
    \begin{tabular}{||c c c c c||}
        \hline
        Tensor Network & Qubits & Avg Dense Flops & Avg SST Flops  & Improvement Factor \\
        \hline\hline
        Concat Rep     & 9      & \num{6.20e+01}  & \num{6.20e+01} & \num{1.000e+00}    \\
        Concat Rep     & 27     & \num{3.20e+02}  & \num{3.12e+02} & \num{1.026e+00}    \\
        Concat Rep     & 81     & \num{1.11e+03}  & \num{1.10e+03} & \num{1.005e+00}    \\
        \hline
        Holographic    & 25     & \num{3.51e+03}  & \num{2.67e+03} & \num{1.316e+00}    \\
        Holographic    & 95     & \num{5.07e+06}  & \num{2.94e+06} & \num{1.722e+00}    \\
        \hline
        RSC            & 9      & \num{2.03e+02}  & \num{1.37e+02} & \num{1.474e+00}    \\
        RSC            & 25     & \num{1.83e+03}  & \num{1.26e+03} & \num{1.452e+00}    \\
        RSC            & 49     & \num{1.37e+04}  & \num{8.30e+03} & \num{1.661e+00}    \\
        \hline
        RSC Tanner     & 9      & \num{3.19e+02}  & \num{3.00e+02} & \num{1.065e+00}    \\
        RSC Tanner     & 25     & \num{5.30e+03}  & \num{3.29e+03} & \num{1.610e+00}    \\
        \hline
        RSC MSP        & 9      & \num{1.30e+03}  & \num{1.19e+03} & \num{1.096e+00}    \\
        RSC MSP        & 25     & \num{8.86e+03}  & \num{6.58e+03} & \num{1.347e+00}    \\
        \hline
        Hamming Tanner & 7      & \num{1.17e+03}  & \num{1.03e+03} & \num{1.136e+00}    \\
        Hamming Tanner & 15     & \num{6.00e+06}  & \num{1.86e+06} & \num{3.224e+00}    \\
        \hline
        Hamming MSP    & 7      & \num{2.54e+03}  & \num{2.00e+03} & \num{1.269e+00}    \\
        Hamming MSP    & 15     & \num{5.48e+09}  & \num{1.15e+06} & \num{4.758e+03}    \\
        \hline
        BB Tanner      & 18     & \num{7.08e+09}  & \num{9.65e+08} & \num{7.336e+00}    \\
        BB Tanner      & 30     & \num{7.03e+12}  & \num{6.20e+12} & \num{1.134e+00}    \\
        \hline
        BB MSP         & 18     & \num{1.23e+13}  & \num{3.73e+08} & \num{3.293e+04}    \\
        BB MSP         & 30     & \num{1.50e+18}  & \num{1.08e+13} & \num{1.389e+05}    \\
        \hline
    \end{tabular}
    \caption{Table showing the average contraction costs found for the \texttt{Hyper-Greedy} method by the two cost functions: dense cost function and our SST cost function. It also shows the improvement factor (dense average cost / SST average cost).}
    \label{table: costs for greedy}
\end{table*}

\begin{table*}
    \begin{tabular}{||c c c c c||}
        \hline
        Tensor Network & Qubits & Avg Dense Flops & Avg SST Flops   & Improvement Factor \\
        \hline\hline
        Concat Rep     & 9      & \num{6.200e+01} & \num{6.200e+01} & \num{1.000e+00}    \\
        Concat Rep     & 27     & \num{3.269e+02} & \num{3.257e+02} & \num{1.004e+00}    \\
        Concat Rep     & 81     & \num{1.242e+03} & \num{1.116e+03} & \num{1.113e+00}    \\
        \hline
        Holographic    & 25     & \num{3.408e+03} & \num{2.705e+03} & \num{1.260e+00}    \\
        Holographic    & 95     & \num{2.585e+06} & \num{2.021e+06} & \num{1.279e+00}    \\
        \hline
        RSC            & 9      & \num{3.360e+02} & \num{1.400e+02} & \num{2.400e+00}    \\
        RSC            & 25     & \num{1.993e+03} & \num{8.762e+02} & \num{2.274e+00}    \\
        RSC            & 49     & \num{1.503e+04} & \num{5.078e+03} & \num{2.961e+00}    \\
        \hline
        RSC Tanner     & 9      & \num{3.30e+02}  & \num{3.25e+02}  & \num{1.013e+00}    \\
        RSC Tanner     & 25     & \num{3.71e+03}  & \num{3.25e+03}  & \num{1.140e+00}    \\
        \hline
        RSC MSP        & 9      & \num{1.38e+03}  & \num{1.32e+03}  & \num{1.049e+00}    \\
        RSC MSP        & 25     & \num{1.28e+04}  & \num{1.02e+04}  & \num{1.258e+00}    \\
        \hline
        Hamming Tanner & 7      & \num{1.27e+03}  & \num{1.15e+03}  & \num{1.099e+00}    \\
        Hamming Tanner & 15     & \num{4.06e+06}  & \num{2.54e+06}  & \num{1.597e+00}    \\
        \hline
        Hamming MSP    & 7      & \num{3.22e+03}  & \num{1.94e+03}  & \num{1.655e+00}    \\
        Hamming MSP    & 15     & \num{9.35e+07}  & \num{3.14e+05 } & \num{2.975e+02}    \\
        \hline
        BB Tanner      & 18     & \num{7.51e+09}  & \num{1.05e+08}  & \num{7.102e+01}    \\
        BB Tanner      & 30     & \num{6.60e+12}  & \num{2.91e+11}  & \num{2.265e+01}    \\
        \hline
        BB MSP         & 18     & \num{5.18e+10}  & \num{3.33e+06}  & \num{1.555e+04}    \\
        \hline
    \end{tabular}
    \caption{Table showing the average contraction costs found for the \texttt{Hyper-Par} method by the two cost functions: dense cost function and our SST cost function. It also shows the improvement factor (dense average cost / SST average cost).}
    \label{table:costs for hyperpar}
\end{table*}
\end{document}